\documentclass{ieeetran}

\IEEEoverridecommandlockouts

\usepackage{graphicx,psfrag}
\usepackage{amsmath,amssymb,amsfonts}
\usepackage{color}
\usepackage{algorithmic}

\newtheorem{theorem}{Theorem}

\newtheorem{proposition}{Proposition}

\newcommand\myrule{\hrule width \columnwidth height .4pt}

\newcommand{\until}[1]{\{1,\dots, #1\}}
\newcommand{\subscr}[2]{#1_{\textup{#2}}}

\newcommand{\setdef}[2]{\{#1 \; : \; #2\}}

\newcommand{\map}[3]{#1: #2 \rightarrow #3}

\newcommand{\integernonnegative}{\ensuremath{\mathbb{Z}}_{\ge 0}}
\newcommand{\real}{\ensuremath{\mathbb{R}}}
\newcommand{\realpositive}{\ensuremath{\real_{>0}}}
\newcommand{\realnonnegative}{\ensuremath{\real}_{\geq0}}

\newcommand{\Z}{\mathbb{Z}}

\newcommand{\R}{\mathbb{R}}

\newcommand{\ie}{{\it i.e.}}
\newcommand{\eps}{\varepsilon}
\newcommand{\E}{\mathcal{E}}    

\newcommand{\neigh}[1]{{\cal N}_{#1}}
\newcommand{\qd}{\texttt{q}}    

\newcommand{\degmax}{\subscr{d}{max}}
\renewcommand{\deg}{d}

\newcommand{\I}{\mathcal{I}}
\newcommand{\ave}{\operatorname{ave}} %
\newcommand{\qave}{\operatorname{qave}} %
\newcommand{\diam}{\operatorname{diam}} %
\newcommand{\sign}{\operatorname{sign}} %

\newcommand{\taumax}{\subscr{\tau}{max}}
\newcommand{\Rmin}{\subscr{R}{min}}

\newcommand{\dst}{\displaystyle}

\def\be{\begin{equation}}
\def\ee{\end{equation}}
\def\ba{\begin{array}}
\def\ea{\end{array}}
\def\eqa{\begin{eqnarray}}
\def\eqe{\end{eqnarray}}

\newcommand{\argmax}{\operatorname{argmax}}

\begin{document}

\title{Robust self-triggered coordination with ternary controllers\thanks{This work is partially supported by the Dutch Organization for Scientific Research (NWO)  under the auspices of the project QUICK (QUantized Information Control for formation Keeping). A preliminary and incomplete account of this work has been submitted for inclusion in the Proceedings of the 3rd IFAC Workshop on Distributed Estimation and Control in Networked Systems.}}

\author{Claudio De Persis
\thanks{
C. De Persis is with ITM, Faculty of Mathematics and Natural Sciences, University of Groningen, Netherlands, and Dipartimento di Ingeneria Informatica Automatica e Gestionale, 
Sapienza Universit\`a di Roma, Italy, (e-mail: c.de.persis@rug.nl).}%
and
Paolo Frasca
\thanks{
P. Frasca is with Dipartimento di Scienze Matematiche, Politecnico di Torino, 10129 Torino, Italy (e-mail: paolo.frasca@polito.it).}}

\maketitle

\begin{abstract}                
This paper regards coordination of networked systems, which is studied in the framework of hybrid dynamical systems.  We design a coordination scheme which combines the use of ternary controllers with a self-triggered communication policy. The communication policy requires  the agents to collect, at each sampling time, relative measurements of their neighbors' states: the collected information is then used to update the control and determine the following sampling time. We prove that the proposed scheme ensures finite-time convergence to a neighborhood of a consensus state. 
%
We then study the robustness of the proposed self-triggered coordination system with respect to skews in the agents' local clocks, to delays, and to limited precision in communication.
Furthermore, we present two significant variations of our scheme. First, we design a time-varying controller which asymptotically drives the system to consensus.
Second, we adapt our framework to a communication model in which an agent  does not poll all its neighbors simultaneously, but single neighbors instead. 
This communication policy actually leads to a self-triggered ``gossip'' coordination system.
\end{abstract}




\section{Introduction}

The key issue in distributed and networked systems resides in ensuring performance with respect to a given control task (e.g. stability, coordination), in spite of possibly severe communication constraints.
In practice, although the system may be naturally described by a continuous-time dynamics,  the control law is only updated at discrete time instants: these can either be pre-specified (time-scheduled control), or be determined by certain events that are triggered depending on the system's behavior.
%
In a networked system, controls and triggering events regarding an agent must only depend on the states (or the outputs) of the agent's neighbors and of the agent itself. Actually, of special interest in distributed systems are self-triggered policies, in which communication and control actions are planned ahead in time, depending to the information currently available at each agent. Indeed, the implementation of an event-based policy, which requires continuous monitoring of a 
triggering condition depending on the state of the agents' neighbors,
may not be suitable to networked applications when sensing and communication resources are critical.

\subsection*{Statement and summary of contributions}

As a main contribution, in this paper we design a new  self-triggered consensus system.
 At each sampling time, a certain subset of ``active'' agents poll their neighbors obtaining relative measurements of the consensus variable of interest: the available information is then used by the active agents to update their controls and compute their next update times. In our system, controls are constrained to belong to $\{-1,0,+1\}$: the assumption of such coarsely quantized controllers is motivated by methodological and opportunity reasons. On one hand, we are interested in demonstrating the effectiveness of ternary controllers for self-triggered coordination. On the other hand, using constrained controllers provides implicit information on the agent dynamics, which can be effectively exploited in designing a self-triggering policy.

Our modeling and design approach naturally leads to a hybrid system which is defined in Section~\ref{sec:model}. Next, in Section~\ref{sect:analysis} we prove, by a Lyapunov analysis, that the hybrid system converges in finite-time to a condition of ``practical consensus'': that is, the solutions are within a neighborhood of the consensus point, and the size of the neighborhood can be made arbitrarily small by decreasing a certain parameter of the 
controller quantizer. 
This parameter, which we denote by $\eps$, 
represents the {\em sensitivity} of the quantizer: the smaller $\eps$, the more the system is demanding in terms of communication resources. We thus identify a trade-off between communication and coordination performance. 
This trade-off is precisely quantified: we provide $\eps$-dependent estimates of the time taken by the solution to reach consensus and of the number of times the agents exchange information.

In self-triggered control, (pre)computation of the sampling times requires a precise knowledge of the system's dynamics. 
Hence, uncertainty in the system can potentially disrupt the correct operation of the control algorithm. Nevertheless we show that the closed-loop system we propose is robust to a variety of uncertainties and disturbances which are relevant in networked systems such as imprecise clock skews, delays and limitations in data rates. Such robustness may be enhanced by introducing a {\em conservativeness} parameter $\alpha$ in the triggering functions which determine the sampling times: the smaller $\alpha$, the shorter are the intervals between sampling times.
The robustness of the control algorithm is studied in Section~\ref{sect:robustness}, by analyzing two extended models, which  include both the conservativeness parameter $\alpha.$

In view of the mentioned need for predictions, it  is also notable that our controllers do not require any information on the network (such as its algebraic connectivity or the number of agents). Furthermore, they only rely on relative measurements: this feature contrasts with other approaches in the literature, which require the knowledge of absolute state information.

As an additional contribution, we show that a suitable time-varying controller, which is designed as a modification of the model introduced in Section~\ref{sec:model}, can asymptotically drive  the system to a consensus state. In this modified version, presented in Section~\ref{sec:asymptotical}, we introduce a time-dependent sensitivity threshold and a time-dependent gain parameter, which both decrease with time. 
In this framework, the time-dependent gain is used to scale the ternary controllers which were used previously. 

In the control scenarios we consider in Sections~\ref{sec:model}-\ref{sec:asymptotical}, every time an agent needs new information, it collects such information from  all its neighbors simultaneously. In Section~\ref{sec.independent}, we show that this simultaneous action is not necessary. We indeed design a self-triggered policy, in which the agents are free to poll their neighbors singularly, and prove for it similar convergence results as before. This system involves variables which are associated to the edges of the graph representing the communication network, and which are updated synchronously by both agents insisting on an active edge. This feature makes the scheme a first example of self-triggered ``gossip'' coordination system.

\subsection*{Literature review}
The reference literature for this paper includes quantized and self-triggered controls for distributed systems. 
Many papers have studied quantization issues in coordination: specifically, ternary (sign) controllers are used to stabilize consensus in~\cite{JC:06b}. In a centralized setting, the use of ternary controllers in connection with quantized communication has been investigated in~\cite{CDP:09}.

Since the seminal work in~\cite{KA-BB:02}, the control community has been interested in investigating event-based and self-triggered control policies. 
In this framework, we note that robustness issues --with respect to parameter uncertainties, delays, and communication losses-- are studied in~\cite{XW-MDL:09}, \cite{MDB-SDG-AD:11},  \cite{HY-PJA:11a} and~\cite{DL-JL:12}.
Relevant papers focusing on networked systems include~\cite{PT:07}, \cite{XW-MDL:11}, \cite{MM-PT:11}, \cite{MM-MC:11}, \cite{CDP-RS-FW:11}, and~\cite{RB-FA:11}.
The work in~\cite{FC-CDP-PF:10a} is also related, as it presents a hybrid coordination dynamics requiring communication only when specific thresholds are met. 

Recent closely related work includes the solution of coordination problems using self-triggered broadcast communication in~\cite{GS-DVD-KHJ:11}. Compared to this reference, the present manuscript proposes a different communication policy, which is based on polling the neighbors upon need, instead than on broadcasting to them. 
An approach which involves polling neighbors has also been considered in the recent paper \cite{DVD-EF-KHJ:12}. Our contribution differs from~\cite{DVD-EF-KHJ:12} in a number of aspects, including the following ones. First, our approach relies on relative measurements and not on absolute ones. Second, in \cite{DVD-EF-KHJ:12} the computation of the next sampling time by an agent requires information not only from the agent's neighbors, but also from the neighbors of the agent's neighbors (\ie~two-hop neighbors). Third, while in \cite{DVD-EF-KHJ:12} zero execution time is allowed (this happens when an agent's local average converges to zero in finite time), in our approach inter-execution times are guaranteed to be bounded away from zero, and the lower bounds are explicitly characterized. 

Self-triggered policies have also been used for deployment of robotic networks in~\cite{CN-JC:12}: in this paper, the authors exploit the knowledge of the speed of the deploying robots in order to design the self-triggering policy. A similar idea features in our present work.

\paragraph*{Notation} 
Notation in this paper is standard. We denote by $\real$, $\realpositive$, $\realnonnegative$ the sets of real, positive, and nonnegative numbers, respectively; by $\integernonnegative$ the set of nonnegative integers.

\section{System definition and main result}\label{sec:model}
\begin{figure*}
\psfrag{Time}{$t$} 
\includegraphics[width=\columnwidth]{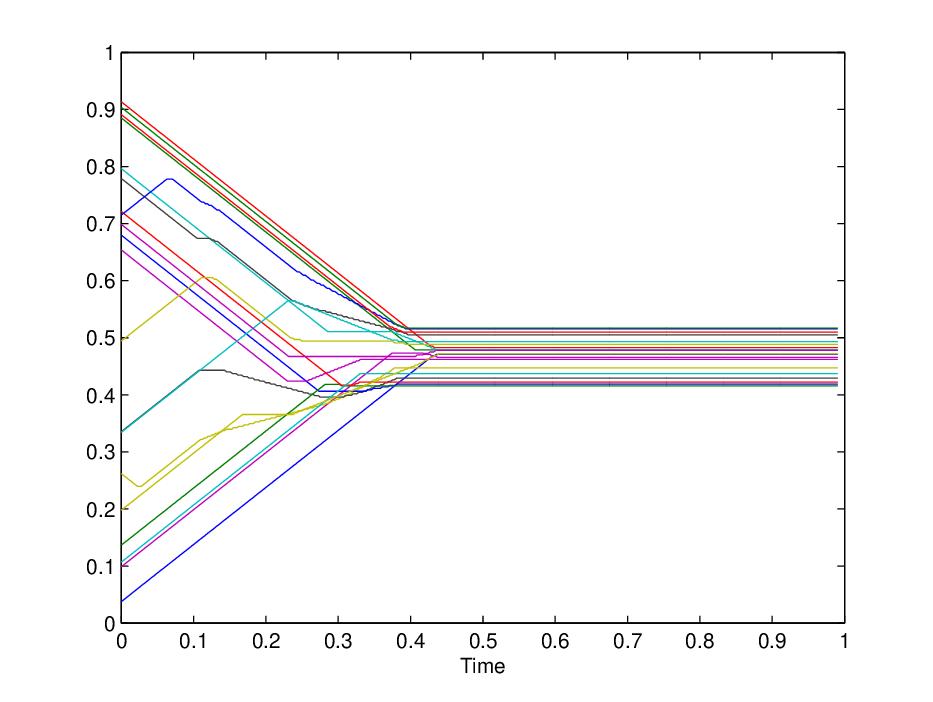}
\includegraphics[width=\columnwidth]{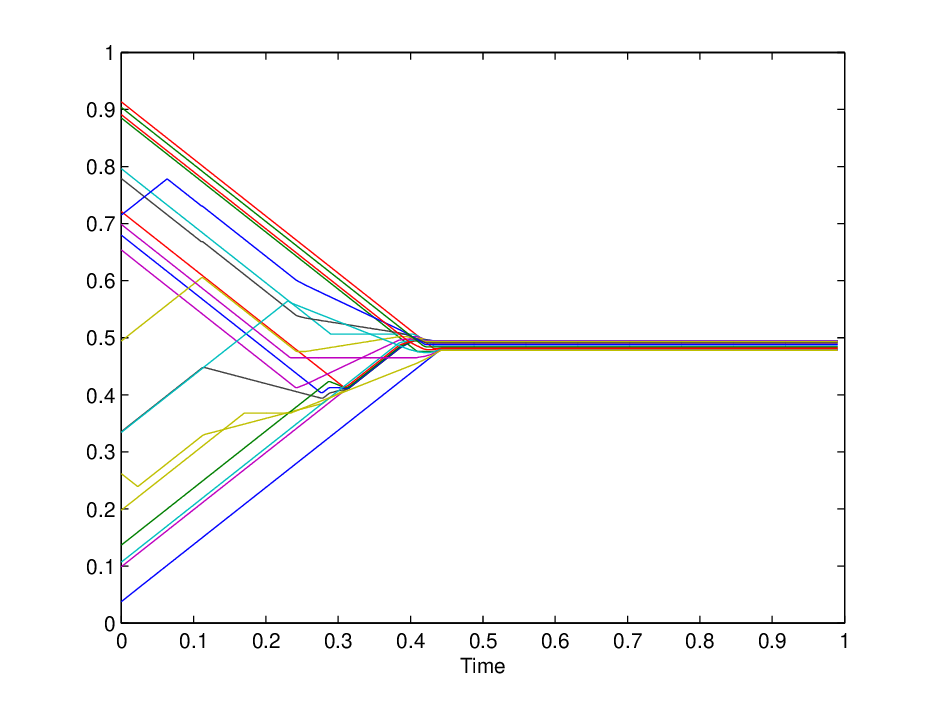}
\caption{Two sample evolutions of the states $x$ in~\eqref{eq:modelloA-cont}-\eqref{eq:modelloA-disc} starting from the same initial condition and on the same graph (a ring graph with $n=20$ nodes). Left plot assumes $\eps=0.01$, right plot assumes $\eps=0.001$.}
\label{fig:simul-modelloA}
\end{figure*}

We assume to have a set of nodes $I=\until{n}$ and an undirected\footnote{We note that this assumption entails  communication in both directions between pairs of connected nodes. However, our communication protocol -- described in Protocol~A 
-- does not require synchronous bidirectional communication, 
} 
graph $G=(I,E)$ with $E$ a set of unordered pairs of nodes, called edges. We denote by $L$ the Laplacian matrix of $G$, which is a symmetric matrix. For each node $i\in I$, we denote by $\neigh{i}$ the set of its neighbors, and by $\deg_i$ its degree, that is, the cardinality of $\neigh{i}.$

We consider the following hybrid dynamics on a triplet of $n$-dimensional variables involving the {\em consensus variable} $x$, the {\em controls} $u$, and the {\em local clock} variables 
$\theta$. All these variables are defined for time $t\ge0$.
Controls are assumed to belong to $\{-1,0,+1\}$. The specific quantizer of choice is $\map{\sign_\eps}{\real}{\{-1,0,+1\}}$, defined according to 
\be\label{eq:signeps} \sign_\eps(z)=\begin{cases} \sign(z) & \text{if}\: |z|\ge\eps\\
0 & \text{otherwise}
\end{cases}\ee
where $\eps>0$ is a {\em sensitivity} parameter.

The system 
$(x,u,\theta)\in\real^{3n}$ 
satisfies the following continuous evolution
\be\label{eq:modelloA-cont} \begin{cases}
\dot x_i=u_i\\
\dot u_i=0\\
\dot \theta_i=-1
\end{cases}\ee
except for every $t$ such that the set 
$\I(\theta,t)=\setdef{i\in I}{\theta_i=0}$ 
is non-empty. At such time instants the system satisfies the following discrete evolution
\be\label{eq:modelloA-disc}
\begin{cases} x_i(t^+)=x_i(t) \quad \forall i\in I \\
u_i(t^+)=
\begin{cases}
\sign_{\eps}\!\left(\ave_i(t)\right) \quad \text{if}\: i\in \I(\theta,t)\\
u_i(t) \quad \text{otherwise}
\end{cases}\\
 \theta_i(t^+)=\begin{cases} f_i(x(t)) \quad \text{if}\: i\in \I(\theta,t)\\
\theta_i(t) \quad \text{otherwise}
\end{cases}\\
\end{cases}
\ee
where for every $i\in I$ the map $\map{f_i}{\real^n}{\realpositive}$ is defined by 
\begin{equation}\label{f_i}
f_i(x)=\begin{cases}
\dst\frac1{4 \deg_i}\,{|\!\sum_{j\in\neigh{i}}(x_j-x_i)|}& \quad \text{if}\: |\dst\sum_{j\in\neigh{i}}(x_j-x_i)|\ge\eps\\
\dst\frac{\eps}{4\deg_i} &\quad \text{otherwise}
\end{cases}
\end{equation}
and for brevity of notation we let
%
\be\label{avei}
\ave_i(t)=\sum_{j\in\neigh{i}}(x_j(t)-x_i(t)).
\ee

The intuition behind the design of the above controller is the following: as we shall verify later,  \eqref{f_i} is such that at each time $t$ and for each $i\in I$, the sign of $u_i(t)$ is consistent with the sign of the ``ideal'' coordination control $\ave_i(t)$, \ie, $u_i(t)\ave_i(t)\ge 0$. This consistency is the key to ensure the desired convergence properties.

It is worth to remark that, although an absolute time variable $t$ is used in the system's definition, and in the analysis which follows, the agents implementing Protocol~A do {\em not} need to be aware of such an absolute time.
Instead, they rely on their local clocks~$\theta_i$. 
Actually, the jump times of each variable $\theta_i$ naturally define a sequence of local switching times, which we denote by $\{t^i_k\}_{k\in \integernonnegative}$. 
Initial conditions can be chosen as $x(0)=\bar x\in\real^n$, 
$u(0)\in \{-1,0,1\}^n$, 
$\theta(0)=0$. With this choice of initial conditions, we note that $\I(0,0)= I$, that is, every agent undergoes a discrete update at the initial time: $t^i_0=0$ for every $i\in I$. We also remark that  inherent in the definition of the discrete evolution (\ref{eq:modelloA-disc}), (\ref{f_i}) is the property that  the period between two consecutive updates of agent $i$'s controller is never smaller than $\frac{\eps}{4d_i}$.

The model~\eqref{eq:modelloA-cont}-\eqref{eq:modelloA-disc} describes the following protocol, which is implemented by each agent $i$ to collect information and compute the control law:
\medskip\myrule
\vspace{.75\smallskipamount}
\noindent\hfill\textbf{Protocol A}\hfill\vspace{.75\smallskipamount}
\myrule\vspace{.75\smallskipamount}
\begin{algorithmic}[1]
\STATE {\bf initialization:} for all $i\in I$, set
$u_i(0)\in \{-1,0,+1\}$ and $\theta_i(0)=0$;
\FORALL{$i\in I$}
\WHILE{$\theta_i(t)>0$} 
\STATE $i$ applies the control $u_i(t)$;
\ENDWHILE
\IF{$\theta_i(t)=0$}
\FORALL{$j\in \neigh{i}$} 
\STATE $i$ polls $j$ and collects the information $x_j(t)-x_i(t)$;
\ENDFOR
\STATE $i$ computes $\ave_i(t)$; 
\STATE $i$ computes $\theta_i(t^+)=f_i(x(t));$
\STATE $i$ computes $u_i(t^+)$ by~\eqref{eq:modelloA-disc};
\ENDIF
\ENDFOR
\end{algorithmic}
\vspace{.5\smallskipamount}\myrule\smallskip

\smallskip

After these remarks, we are ready to state our first convergence result:
\begin{theorem}[Practical consensus]\label{thm:modelloA-convergence}
For every initial condition $\bar x$, let $x(t)$ be the solution to~\eqref{eq:modelloA-cont}-\eqref{eq:modelloA-disc} such that $x(0)=\bar x$. 
Then $x(t)$ converges in finite time to a point $x^*$ belonging to the 
set 
\be\label{set.E}
\E=
\setdef{x\in \R^n}{|\sum_{j\in \neigh{i}} (x_j-x_i)|< \eps \;\forall\,i\in I}.
\ee
\end{theorem}
This result can be seen as a ``practical consensus'' result, as the size of the consensus error can be be made as small as needed by choosing $\eps$.
Moreover, we can estimate the time and communication costs of the system, as follows:
\begin{proposition}[Time and communication costs]\label{prop:costs}
Let $x(\cdot)$ be the solution to system~\eqref{eq:modelloA-cont}-\eqref{eq:modelloA-disc}. Define the {\em time cost} $T=\inf\setdef{t\ge0}{x(t)\in \E}$
and the {\em communication cost}
$C=\max_{i\in I} \max\setdef{k}{t_k^i\le T}.$
Then, 
$$ T\le \frac{2(1+\degmax)}\eps \sum_{\{i,j\}\in E}(\bar x_i-\bar x_j)^2$$
and
$$ C\le \frac{8\degmax(1+\degmax)}{\eps^2}
\sum_{\{i,j\}\in E}(\bar x_i-\bar x_j)^2,$$
where $\bar x\in\real^n$ is the initial condition.
\end{proposition}

\medskip

Since each polling action involves polling at most $\degmax$ neighbors, we also conclude that the total number of messages to be exchanged in the whole network in order to achieve (practical) consensus is not larger than 
$$\frac{8\degmax^2(1+\degmax) n }{\eps^2}
\sum_{\{i,j\}\in E}(\bar x_i-\bar x_j)^2.$$ 

\begin{figure}
{
\psfrag{x}[][][1][-90]{$x$} 
\psfrag{data1}[][][1][0]{$x_1$} 
\psfrag{data2}[][][1][0]{$x_2$} 
\psfrag{data3}[][][1][0]{$x_3$} 
\psfrag{data4}[][][1][0]{$x_4$} 
\psfrag{data5}[][][1][0]{$x_5$} 
\psfrag{u1}[][][1][-90]{ $u_1$} 
\psfrag{u2}[][][1][-90]{ $u_2$} 
\psfrag{u3}[][][1][-90]{ $u_3$} 
\psfrag{u4}[][][1][-90]{$u_4$} 
\psfrag{u5}[][][1][-90]{$u_5$} 
\psfrag{Time}{$t$} 
\includegraphics[width=\columnwidth]{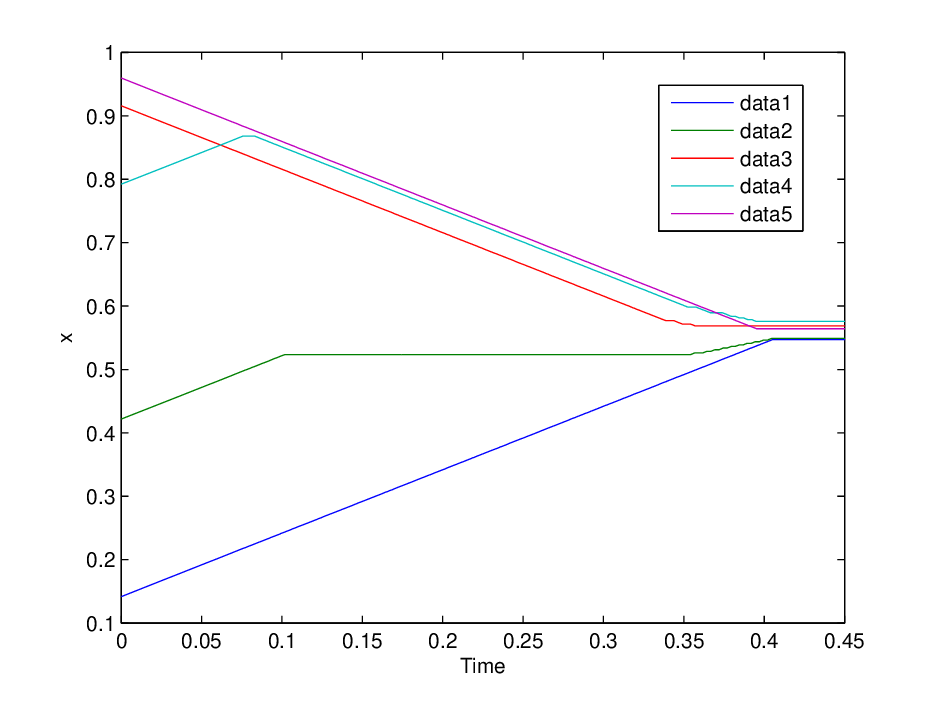}
\includegraphics[width=\columnwidth]{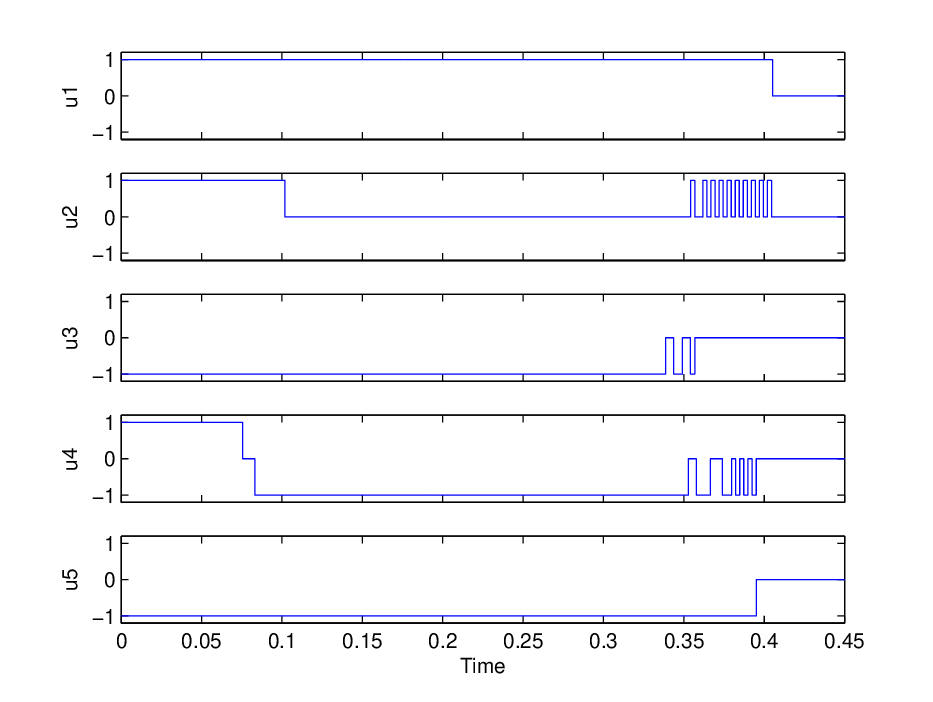}
}
%
\caption{Sample evolutions of states $x$ and corresponding controls $u$ in~\eqref{eq:modelloA-cont}-\eqref{eq:modelloA-disc} on a ring with $n=5$ nodes, $\eps=0.02$.}
\label{fig:simul-modelloA-controls}
\end{figure}

Our theoretical results suggest that, by choosing the sensitivity $\eps$, we are trading between precision and cost, both in terms of time and of communication effort. However, simulations indicate that the role of $\eps$ in controlling the speed of convergence is limited, as long as $x(t)$ is far from $\E$. Before approaching the limit set, solutions are qualitatively similar to the solutions of consensus dynamics with controls in $\{-1,+1\}$: this remark is confirmed if we compare Fig.~\ref{fig:simul-modelloA} with Fig.~1~(rightmost) in~\cite{JC:06b}. Consistently, Fig.~\ref{fig:simul-modelloA-controls} demonstrates that the state trajectories ``brake'', and the controls switch between zero and non-zero, as the states approach the region of convergence. Once this is reached (in finite time), the controls stop switching and remain constantly to zero, as the analysis in the next section shows.


\subsection{Convergence analysis}\label{sect:analysis}
This subsection is devoted to the proofs of Theorem~\ref{thm:modelloA-convergence} and Proposition~\ref{prop:costs}.
\begin{IEEEproof}[Proof of Theorem~\ref{thm:modelloA-convergence}]
First of all, we recall that 
\be\label{eq:st}
t_{k+1}^i=t_k^i +
\left\{\ba{ccc}
\dst\frac{|\ave_i(t_k^i)|}{4d_{i}} & {\rm if}  &
|\ave_i(x(t_k^i ))|\ge \eps\\[4mm]
\dst\frac{\eps}{4d_{i}} 
& {\rm if} & |\ave_i(x(t_k^i ))|< \eps.
\ea
\right.
\ee
Then, we immediately argue that, for every $i\in I$, the sequence of local switching times $\{t^i_k\}_{k\in \integernonnegative}$ has the following ``dwell time'' property:
for every $k\ge 0$,
\be\label{eq:DeltaT-bound}t^i_{k+1}-t^i_{k}\ge \frac{\eps}{4 \degmax}.\ee   
Inequality~\eqref{eq:DeltaT-bound} implies that there exists a positive dwell time between subsequent switches and this fact in turn implies that for each initial condition,
\eqref{eq:modelloA-cont} has a piecewise constant right-hand side.
Hence the system has a unique solution $x(\cdot)$, which is an absolutely continuous function of its time argument. 
Furthermore, solutions are bounded, since one can show that for all $t>0$ it holds that $\min_i x_i(0)\le \min_i x_i(t)$ 
and $\max_i x_i(0)\ge \max_i x_i(t).$
We are interested in studying the convergence properties of such solutions.
For every $t\ge0$, we let $$V(t)=\frac{1}{2}{x^T(t) L x(t)}.$$ We note that $V(t)\ge 0$ and we consider the evolution of $\dot V(t)$ along the solution. 
Since $L$ is symmetric, and letting $t_k^i=\max\setdef{t_h^i}{t_h^i<t, j\in \integernonnegative}$, we have
\begin{align*}
\dot V(t)=&\,x^T(t) L u(t) =u^T(t) L x(t)
\\= &-\dst\sum_{i=1}^{n} 
\left(
\dst\sum_{j\in \neigh{i}} (x_j(t)-x_i(t)) 
\right)
\sign_\eps\!\left({\ave_{i} (t_k^i)}\right) 
\\
=& -\!\!\!\!\!\!
\dst\sum_{i:|\ave_{i} (t_k^i)|\ge \eps} 
\!\!\!
\ave_i(t)\,
\sign_\eps\!\left({\ave_{i} (t_k^i)}
\right).
\end{align*}
Using Eq.~\eqref{eq:st} we observe that,
for $t\in [t_{k}^i, t_{k+1}^i]$, if $\ave_i(t^i_k)\ge\eps$, then
\begin{align}
\nonumber
\dst\sum_{j\in \neigh{i}} (x_j(t)-x_i(t)) 
\ge & 
\dst\sum_{j\in \neigh{i}} (x_j(t_k^i)-x_i(t_k^i)) 
-2 d_i (t-t_k^i)\\
\label{eq:ave(t)-below}
\ge & 
\dst\frac{\ave_i(t_k^i)}{2}
\end{align}
Similarly, if $\ave_i(t^i_k)\le-\eps$, then
\[
\dst\sum_{j\in \neigh{i}} (x_j(t)-x_i(t)) 
\le - \dst\frac{\ave_i(t_k^i)}{2}. 
\]

These inequalities imply that, if  $|\ave_{i} (t_k^i)|\ge \eps$, 
then 
$\sum_{j\in \neigh{i}} (x_j(t)-x_i(t)) $ preserves the sign during continuous flow
by continuity of $x(t)$, and consequently 
\begin{align}
\nonumber
\ave_i(t)
\sign_\eps (\ave_{i} (t_k^i)) 
&=\ave_i(t)
\sign(\dst\sum_{j\in \neigh{i}} (x_j(t)-x_i(t)) ) 
\\
\label{eq:abs-no-abs}
&=\left|
\ave_i(t)
\right|.
\end{align}
Moreover,
\begin{align}
\nonumber
\left|
\dst\sum_{j\in \neigh{i}} (x_j(t)-x_i(t)) 
\right|
\ge & 
\left|
\dst\sum_{j\in \neigh{i}} (x_j(t_k^i)-x_i(t_k^i)) 
\right|
-2 d_i (t-t_k^i)\\
\label{eq:ave(t)-below}
\ge & 
\dst\frac{|\ave_i(t_k^i)|}{2}
\end{align}
Hence, using Equations~\eqref{eq:abs-no-abs} and~\eqref{eq:ave(t)-below} we deduce 
\be\label{eq:dotV-bound}
\dot V(t)\le -\dst\sum_{i:|\ave_{i} (t_k^i)|\ge \eps} \frac{|\ave_i(t_k^i)|}{2}\le  -\dst\sum_{i:|\ave_{i} (t_k^i)|\ge \eps} \frac{\eps}{2}.
\ee
This inequality implies there exists a finite time $\bar t$ such that $|\ave_i(t_k^i)|<\eps$ for all $i\in  I$ and all $k$ such that  $t_k^i\ge\bar t$. 
Indeed, otherwise there would be an infinite number of time intervals whose length is bounded away from zero and on which $\dot V(t)\le -\frac{\eps}2$, contradicting the positivity of $V$.
For all $i\in I$, let ${\bar k_i}=\min\setdef{k\ge0}{t_k^i\ge\bar t\,}$
and define 
$$\hat t=\inf\setdef{t\ge0}{t>t_{\bar k_i}^i \;\text{for all}\; i\in I}.$$ Note that $\hat t\ge \bar t$ and thus $|\ave_i(t_k^i)|<\eps$ if $t_k^i\ge \hat t $.
Moreover, by definition of $\hat t$,  for $t\ge \hat t$ and for all $i=1,2,\ldots, n$, the controls $u_i(t)$   are zero  and the states $x_i(t)$ are constant and such that 
$|\ave_i(t)|<\eps$ for all $i\in I$. 

We conclude that there exists a point $x^*\in \real^n$ such that $x(t)=x^*$ for $t\ge \hat t$, and 
$$x^*\in\{x\in \R^n: |\sum_{j\in \neigh{i}} (x_j-x_i)|<
\eps, \forall i\in  I\}. 
$$
\end{IEEEproof}


The above proof shows that convergence is reached in finite time: obtaining an estimate of this convergence time requires a deeper look into the dynamics of the system.
\begin{IEEEproof}[Proof of Proposition~\ref{prop:costs}]
In order to prove Proposition~\ref{prop:costs}, we recall that Eq.~\eqref{eq:dotV-bound} implies that for every $t\ge0$,
$$\dot V(t)\le -\dst\sum_{i:|\ave_{i} (t_k^i)|\ge \eps} \frac{\eps}{2}.$$
We want to use this fact to estimate the time taken by $x(t)$ to reach the set $\E$.
First of all, note that if $u(t)\neq 0$ for all $t<T$, then the set $\{i:|\ave_{i} (t_k^i)|\ge \eps\}$ is not empty and we can argue that the Lyapunov function decreases by at least $\eps/2$ per time unit, until convergence is reached. 

Let us then consider the more interesting case in which there exists $t'< T$ such that $u(t')=0$.
For all $i\in I$, define $k^\star_i=\max\setdef{h}{t^i_h\le t'}$, and 
consider $$t^\star=\inf\setdef{t\ge0}{t>t^i_{k^\star_i} \;\forall \,i\in I}.$$ Clearly $u(t^\star)=0$ and $t^i_{k^\star_i}\le t^\star\le t' \le t^i_{k^\star_i+1}$ for all $i\in I$.
Note that for $u(t')$ to be zero, necessarily $|\ave_{i} (t_{k^\star_i}^i)|< \eps$, and then $t_{k^\star_i+1}^i-t_{k^\star_i}^i=\frac\eps{4 \deg_i}$ for all $i\in I$.
If $|\ave_{i} (t_{k^\star_i+1}^i)|<\eps$ for all $i\in I$ as well, then we can see that $u(t)=0$ for all $t\ge t^\star$, implying that convergence is reached and $T=t^\star
\le t',$ contradiction.

It must then exist\footnote{We remark that the existence of such $j$ is permitted because, although $u(t)=0$ when $t\in (t^\star,\min_i t_{k^\star_i+1}^i)$, actually $u(t)$ needs not to be zero for $t\in (\min_i t_{k^\star_i}^i, t^\star).$} 
$j\in I$ such that $|\ave_{j} (t_{k^\star_j+1}^j)|\ge 
\eps$. Note that $t_{k^\star_j+1}^j-t^\star\le \frac{\eps}{4 \deg_j}\le \frac\eps4$, whereas 
$\dot V(t)\le -\frac\eps2$ for $t\in (t_{k^\star_j+1}^j,t_{k^\star_j+1}^j+\frac{\eps}{4 \deg_j}).$ 
The discussion above yields the following conclusion.
Before convergence is reached, controls may possibly be zero and the set $\{i:|\ave_{i} (\cdot)|\ge \eps\}$ may be empty: however, this condition may only persist for a duration smaller than $\frac{\eps}{4}$, after which the set $\{i:|\ave_{i} (\cdot)|\ge \eps\}$ is not empty for a time not shorter than $ \frac{\eps}{4 \degmax}$.
Consequently, we argue that every $\frac{\eps}{4}(1+\frac{1}{\degmax})$ units of time, $V(t)$ decreases by at least $\frac{\eps}{2}\cdot  \frac{\eps}{4 \degmax}$. Hence, if 
\[
T> \frac{V(0)}{\frac{\eps}{2}\cdot  \frac{\eps}{4 \degmax}}\cdot \frac{\eps}{4}(1+\frac{1}{\degmax})= \dst\frac{V(0)}{\frac\eps2 \frac1{1+\degmax}},
\]
then the Lyapunov function would become negative, which is a contradiction. This implies that within $ \dst T\le \dst\frac{V(0)}{\frac\eps2 \frac1{1+\degmax}}$ units of time, the system must converge to the set of  states (\ref{set.E}) 
where $V(t)$ is constant. 
Moreover, Eq.~\eqref{eq:DeltaT-bound} implies that the number of communication  events involving any agent $i$ is not larger than
$$ \frac{\frac{2 (1+\degmax)}\eps
V(0)}{\frac{\eps}{4 \degmax}}=\frac{8 \degmax (1+\degmax)}{\eps^2}V(0).$$

The thesis follows if we recall that $\bar x$ is the initial condition and that $V(0)=\sum_{\{i,j\}\in E}(\bar x_i-\bar x_j)^2$.
\end{IEEEproof}

\section{Robustness}\label{sect:robustness}

In this section we discuss the robustness of Protocol~A 
to some typical non-idealities which can occur in its implementation. We consider the issues of clock skew, delays, and limited precision of data: 
while these are not the only issues which can arise,  
we believe they are the most significant to our exposition, which regards networked problems. For simplicity of presentation, we do not study these three issues together: we consider clock skews first in combination with delays, and then with quantization. A model including all three issues can be studied using the same tools.

The key idea to quantify the robustness properties, which are inherent to Protocol~A, involves introducing a design parameter $\alpha$ which represents how conservative the agents are when planning their next sampling time. By proving convergence conditions for such extended model, we shall show that, provided the design parameters $\eps$ and $\alpha$ are properly chosen, our protocol can always be made robust to quantization errors, clock rate variabilities, and delays. The analysis reveals natural trade-offs between robustness and accuracy performance.


\subsection{Clock skews and delays}
In this section, we discuss the intrinsic robustness of Protocol~A against model uncertainties in local clock specifications, combined with communication and actuation delays. To this goal, we extend the protocol to include such delays and clock rate variabilities. We thus generalize system~\eqref{eq:modelloA-cont}-\eqref{eq:modelloA-disc} by considering the system
$(x,u,\theta)\in\real^{3n}$ 
satisfying the continuous evolution
\be\label{eq:modelloA-cont-delay} \begin{cases}
\dot x_i=u_i\\
\dot u_i=0\\
\dot \theta_i=-R_i
\end{cases}\ee
where $R_i>0$ is the rate of the local clock at agent $i$,
and the discrete evolution defined as follows. 
Let the set of {\em switching} agents be defined as $\I(\theta,t)=\setdef{i\in I}{\theta_i(t)=0}.$
Each agent $i\in \I(\theta,t)$ polls its neighbors at time $t$: since implementing communication and actuation entails a nonnegative delay 
$\tau_i(t),$ each switching agent $i\in \I(\theta,t)$ undergoes the following update at time $\bar t=t+\tau_i(t)$:
%
\be\label{eq:modelloA-disc-delay}
\begin{cases} x_i(\bar t^+)=x_i(\bar t) \quad \forall i\in I \\
u_i(\bar t^+)=
\begin{cases}
\sign_{\eps}\!\left(\ave_i(t)\right) \quad \text{if}\: i\in \I(\theta,t)\\
u_i(\bar t) \quad \text{otherwise}
\end{cases}\\
 \theta_i(\bar t^+)=\begin{cases} f^\alpha_i(x(t)) \quad \text{if}\: i\in \I(\theta,t)\\
\theta_i(\bar t) \quad \text{otherwise}
\end{cases}\\
\end{cases}
\ee
where $\ave_i(t)$ is defined as in (\ref{avei}), and 
for every $i\in I$ the map $\map{f^\alpha_i}{\real^n}{\realpositive}$ is defined by 
$$
f_i(x)=\begin{cases}
\dst\frac\alpha{2 \deg_i}\,{|\!\sum_{j\in\neigh{i}}(x_j-x_i)|}& \quad \text{if}\: |\dst\sum_{j\in\neigh{i}}(x_j-x_i)|\ge\eps\\
\dst\frac{\alpha\,\eps}{2\deg_i} &\quad \text{otherwise}
\end{cases}
$$
%
where  $\alpha>0$ is a design parameter. 
The initial conditions are chosen as before, namely 
$x(0)=\bar x\in\real^n$, 
$u(0)\in \{-1,0,1\}^n$, 
$\theta(0)=0$. 
Note that system~\eqref{eq:modelloA-cont}-\eqref{eq:modelloA-disc} is a special case of the above definition, assuming $\alpha=\frac12$, $R_i=1$, and $\tau_i=0$ for all $i\in I$. 
We are now ready to state our robustness result.
\begin{proposition}[Clock skew \& delay robustness]\label{prop:delay-robustness}
Consider system~\eqref{eq:modelloA-cont-delay}-\eqref{eq:modelloA-disc-delay} and assume that $R_i\ge \Rmin>0$ and $\tau_i(\cdot)\le \taumax$ for all $i\in I$. If $\eps>4 \degmax \taumax$ and 
$$ \alpha<\frac{\eps-4 \degmax \taumax}{\eps}{\Rmin},$$
 then $x(t)$ converges to a point in the set $\E$ defined in (\ref{set.E}) in finite time.
\end{proposition}
\begin{IEEEproof}
Similarly to what we did in the previous section, we consider, for every $i\in I$, two sequences: the sequence $\setdef{t^i_h}{h\in\integernonnegative}$ of times at which the agent $i$ polls its neighbors, and the sequence $\setdef{s^i_h}{h\in\integernonnegative}$ of times at which the agent $i$ updates its control, 
with $t^i_0=0$.
Note that
$s^i_h=t^i_h+\tau_i(t^i_h)$, and 
\begin{align*}
t^i_{h+1}=&s^i_h+\frac1{R_i}f^\alpha_i(x(t^i_h))=t^i_h+\tau_i(t^i_h)+\frac1{R_i}f^\alpha_i(x(t^i_h))\\\ge& t^i_h+\frac{\alpha \eps}{2 d_i R_i}\\
s^i_{h+1}=&t^i_{h+1}+\tau_i(t^i_{h+1})=s^i_h+\frac1{R_i}f^\alpha_i(x(t^i_h))+\tau_i(t^i_{h+1})\\\ge& s^i_h+\frac{\alpha \eps}{2 d_i R_i}.
\end{align*}
These inequalities ensure that solutions are well defined.

For all $t\ge0$ and $i\in I$, let $s^i_k=\max\setdef{s^i_h}{s^i_h<t}$ and $t^i_k=\max\setdef{t^i_h}{t^i_h<s^i_k}.$
With these definitions, we have
$$ t-t^{i}_k\le \tau_i(t^i_k) +\frac1{R_i}{f_i^\alpha(x(t^i_k))}
+
\tau_i(t_{k+1}^i).
$$
Using this inequality we observe that, if  $\ave_i(t_k^i)\ge \eps$, then 
\begin{align}
\nonumber
\ave_i(t)
\ge & 
\ave_i(t_k^i) -2 d_i (t-t_k^i)\\
\nonumber
\ge & 
\ave_i(t_k^i) -2 d_i 
\big( \frac{\alpha |\ave_i(t_k^i)|}{2 d_i R_i}+
\tau_i(t_{k}^i)+\tau_i(t_{k+1}^i)
\big) \\
\nonumber
\ge & 
\ave_i(t_k^i)\left(1 -\frac{\alpha}{\Rmin}\right) - 4 \degmax \taumax >0.
\label{eq:ave(t)-below}
\end{align}

We also let $V(t)=\frac{1}{2}{x^T(t) L x(t)}$ for every $t\ge0$, and we consider the evolution of $\dot V(t)$ along the solution. 
We then have
\begin{align*}
\dot V(t)=&\,x^T(t) L u(t) \\
= &-\dst\sum_{i=1}^{n} 
\left(
\dst\sum_{j\in \neigh{i}} (x_j(t)-x_i(t)) 
\right)
u_i({s^i_k}^+) 
\\
= &-\dst\sum_{i=1}^{n} 
\ave_i(t)
\sign_\eps\!\left({\ave_{i} (t_k^i)}\right) 
\\
=& -\!\!\!\!\!\!
\dst\sum_{i:|\ave_{i} (t_k^i)|\ge \eps} 
\!\!\!
\ave_i(t)
\sign_\eps\!\left({\ave_{i} (t_k^i)}
\right).
\end{align*}

From here on, the same reasoning as in the proof of Theorem~\ref{thm:modelloA-convergence} can be applied to show that 
$$
\dot V(t)\le  -\dst\sum_{i:|\ave_{i} (t_k^i)|\ge \eps}\left( \eps \big(1-\frac\alpha{\Rmin}\big)- 4 \degmax \taumax\right) .
$$
From this inequality, a similar Lyapunov argument as in the proof of Theorem~\ref{thm:modelloA-convergence} implies the desired convergence property.
\end{IEEEproof}

We note that, according to Proposition~\ref{prop:delay-robustness}, any (bounded) delay can be tolerated, but entails a proportionally large loss in the achievable precision.   

\subsection{Clock skews and quantized information}

A variation of the control scenario considered so far considers the possibility in which when an agent polls its neighbors it receives quantized information.
This scenario can raise when the agent is endowed with a sensor which provides coarse (quantized) measurements of the neighbors' states. A different scenario is when not all the agents are endowed with sensors able to measure the relative distance from their neighbors. These agents must then receive the information in a quantized form from their neighbors via a digital communication channel. 
We adopt for our analysis a standard uniform quantizer, defined as 
\[
\qd(x)=\Delta\left\lfloor \frac{x}{\Delta}+\frac{1}{2}\right\rfloor
\]
where $\Delta>0$ is a parameter inversely proportional  to the precision of the quantizer.  

To take into account the presence of quantized measurements, model~\eqref{eq:modelloA-cont}-\eqref{eq:modelloA-disc} is modified as follows. The continuous evolution obeys the equations 
\be\label{eq:modelloA-cont-quant} \begin{cases}
\dot x_i=u_i\\
\dot u_i=0\\
\dot \theta_i=-R_i
\end{cases}\ee
where $R_i>0$ are the local clock rates.
At every $t$ such that the set $\I(\theta,t)=\setdef{i\in I}{\theta_i=0}$ 
is non-empty, the system instead satisfies the following discrete evolution:
\be\label{eq:modelloA-disc-quant}
\begin{cases} x_i(t^+)=x_i(t) \quad \forall i\in I \\
u_i(t^+)=
\begin{cases}
\sign_{\eps}\left(\qave_i(t)\right) \quad \text{if}\: i\in \I(\theta,t)\\
u_i(t) \quad \text{otherwise}
\end{cases}\\
 \theta_i(t^+)=\begin{cases} f^\alpha_i(x(t)) \quad \text{if}\: i\in \I(\theta,t)\\
\theta_i(t) \quad \text{otherwise}
\end{cases}\\
\end{cases}
\ee
where for every $\alpha>0$ and $i\in I$ the map $\map{f^\alpha_i}{\real^n}{\realpositive}$ is defined by 
$$
f_i(x)=\begin{cases}
\dst\frac\alpha{2 \deg_i}\,{|\!\sum_{j\in\neigh{i}}\qd(x_j-x_i)|}& \quad \text{if}\: |\dst\sum_{j\in\neigh{i}}\qd(x_j-x_i)|\ge\eps\\
\dst\frac{\alpha\,\eps}{2\deg_i} &\quad \text{otherwise}
\end{cases}
$$
and we have used the notation $$\qave_i(t)=\sum_{j\in\neigh{i}}\qd(x_j(t)-x_i(t)).$$ 
We are now ready to state a second robustness result.
\begin{proposition}[Clock skew \& quantization robustness]\label{prop:quant-robustness}
Consider system~\eqref{eq:modelloA-cont-quant}-\eqref{eq:modelloA-disc-quant} and assume that $R_i\ge \Rmin>0$ for all $i\in I$. If $\eps>\frac12\degmax \Delta$ and 
$$ \alpha<\frac{2\eps-\degmax \Delta}{2\eps}\Rmin,$$
then $x(t)$ converges in finite time to a point in 
\[\E_2=
\{x\in \R^n\,:\, |\sum_{j\in \neigh{i}} (x_j-x_i)|< 2\eps\}. 
\]
\end{proposition}
\begin{IEEEproof}
Similarly to the previous protocols,  this algorithm ensures a guaranteed minimum inter-sampling time given by $\frac{\alpha}{2\degmax}\eps$. Hence, the solutions to the system are well-defined and unique. Along these solutions, the Lyapunov function $V=\frac{1}{2}x^TLx$ satisfies
$$
\dot V(t)=   -\!\!\!\!\!\!
\dst\sum_{i:|\qave_{i} (t_k^i)|\ge \eps} \!\!
\ave_i(t)\,\,
\sign_{\eps}({\qave_{i} (t_k^i)}),$$
where as before $t_k^i$ denotes the largest time at which agent $i$ polls its neighbors before time $t$. 
Observe that for all $t$,
\be\label{eq:compare-qave0}
\ave_i (t)-\frac{\Delta}{2}\deg_i\le \qave_i (t) \le \ave_i (t)+\frac{\Delta}{2}\deg_i.
\ee
and also
\be\label{eq:compare-qave}|\ave_i (t)|-\frac{\Delta}{2}\deg_i\le |\qave_i (t)|\le |\ave_i (t)|+\frac{\Delta}{2}\deg_i.\ee
For $t\in [t^{i}_k, t^{i}_{k+1}]$ and $\qave_i(t_k^i)\ge \eps$
$$t-t^{i}_k\le \frac{\alpha}{2 d_i}{\qave_i(t^i_k)} \frac1{R_i}.$$
Using this fact and (\ref{eq:compare-qave0}), we argue that, if  $\qave_i(t_k^i)\ge \eps$, then
\begin{align*}
\ave_i(t)
\ge & 
\ave_i(t_k^i) -2 d_i (t-t_k^i)\\
\ge & 
\ave_i(t_k^i) -2 d_i \left( \frac{\alpha}{2 d_i}\qave_i(t_k^i)\frac1{R_i}\right) \\
\ge & 
\qave_i(t_k^i) - \frac12 d_i \Delta - \frac\alpha{R_i} \qave_i(t_k^i) \\
\ge & 
\left(1 -\frac{\alpha}{R_{i}}\right) \eps - \frac12 \deg_i \Delta\\
\ge & 
\left(1 -\frac{\alpha}{\Rmin}\right) \eps - \frac12 \degmax \Delta
\end{align*}
An analogous inequality holds in the case $\qave_i(t_k^i)\le -\eps$.  
Using the inequalities above, arguments similar to those  in the proof of Theorem~\ref{thm:modelloA-convergence} lead to 
$$
\dot V(t)\le  -\dst\sum_{i:|\qave_{i} (t_k^i)|\ge \eps} \left(\big (1 -\frac{\alpha}{\Rmin}\big) \eps - \frac12 \degmax \Delta\right),
$$
and ultimately to the convergence in finite time to the set such that 
$$|\sum_{j\in\neigh{i}}\qd(x_j(t)-x_i(t))|<\eps. $$
The result thus follows from~\eqref{eq:compare-qave} and the condition  on $\eps$.
\end{IEEEproof}

We conclude from Proposition~\ref{prop:quant-robustness} that the system is robust to quantized communication, and the achievable precision is proportional to the precision of the quantizer. 


\section{Asymptotical consensus}\label{sec:asymptotical}
In this section we propose a modification of system~\eqref{eq:modelloA-cont}-\eqref{eq:modelloA-disc}, which drives the system to asymptotical consensus.
The key idea involves decreasing the sensitivity threshold with time and concurrently introducing a {\em time-varying decreasing gain} in the control loop.

Let $\map{\eps}{\realnonnegative}{\realpositive}$ and $\map{\gamma}{\realnonnegative}{\realpositive}$ be non-increasing functions such that  $$\lim_{t\to+\infty} \eps(t)=\lim_{t\to+\infty} \gamma(t)=0.$$
We consider the system $(x,u,\tau)\in\real^{3n}$ which satisfies the following continuous evolution%
\footnote{Note that in this case the agents need to evaluate $\gamma$ and $\eps$ as functions of $t$. Hence absolute time is assumed to be known to the agents in this section. For this reason, the robustness properties of~\eqref{eq:modelloB-cont}-\eqref{eq:modelloB-disc} do not trivially follow from the analysis in Section~\ref{sect:robustness}: a detailed study is left to future work.}
\be\label{eq:modelloB-cont} \begin{cases}
\dot x_i=\gamma\, u_i\\
\dot u_i=0\\
\dot \theta_i=-1
\end{cases}\ee
except for every $t$ such that the set $\I(\theta,t)=\setdef{i\in  I}{\theta_i=0}$ is non-empty. 
At such time instants the system satisfies the following discrete evolution
\be\label{eq:modelloB-disc} \begin{cases} x_i(t^+)=x_i(t) \quad \forall i\in  I\\
u_i(t^+)=
\begin{cases}
\sign_{\eps(t)}\left(\ave_i(t)\right) \quad \text{if}\: i\in \I(\theta,t)\\
u_i(t) \quad \text{otherwise}
\end{cases}\\
 \theta_i(t^+)=\begin{cases}
 \frac1{\gamma(t)}f_i(x(t)) \quad \text{if}\: i\in \I(\theta,t)\\
\theta_i(t) \quad \text{otherwise}
\end{cases}\\
\end{cases}\ee
where for every $i\in I$ the maps $\ave_i(t)$ and $f_i(x)$ are the same maps defined earlier in the paper. We also adopt the same initial conditions as before, namely $x(0)=\bar x\in\real^n$, 
$u(0)\in \{-1,0,+1\}$,
$\theta(0)=0$. As a result $t_k^i=0$ for every $i\in \{1, 2, \ldots, n\}$.

The corresponding protocol is the following:
\medskip\myrule
\vspace{.75\smallskipamount}
\noindent\hfill\textbf{Protocol B}\hfill\vspace{.75\smallskipamount}
\myrule\vspace{.75\smallskipamount}
\begin{algorithmic}[1]
\STATE {\bf initialization:} for all $i\in I$, set $u_i(0)\in \{-1,0,+1\}$ and $\theta_i(0)=0$;
\FORALL{$i\in I$}
\WHILE{$\theta_i(t)>0$} 
\STATE $i$ applies the control $u_i(t)$;
\ENDWHILE
\IF{$\theta_i(t)=0$}
\FORALL{$j\in \neigh{i}$} 
\STATE $i$ polls $j$ and collects the information $x_j(t)-x_i(t)$;
\ENDFOR
\STATE $i$ computes $\ave_i(t)$; 
\STATE $i$ computes $\theta_i(t^+)=\frac{1}{\gamma(t)}f_i(x(t));$
\STATE $i$ computes $u_i(t^+)$ by~\eqref{eq:modelloB-disc};
\ENDIF
\ENDFOR
\end{algorithmic}
\vspace{.5\smallskipamount}\myrule\smallskip

 In this new protocol we let the parameter $\eps$,  which -- as established in the previous sections -- gives a measure of the size of the region of practical convergence, to be time-varying and converging to zero. The obvious underlying rationale is that if the size of the convergence region goes to zero as time elapses, one might be able to establish asymptotical convergence rather than practical. However, letting  $\eps$ go to zero does not suffice and may induce agents to poll their neighbors infinitely  often in a finite interval of time (Zeno phenomenon). To prevent this occurrence we slow down both the process of requesting information to the neighbors and the velocity of the system. The former is achieved via a factor $\frac{1}{\gamma(t)}$ multiplying the map $f_i(x)$, the latter via the factor $\gamma(t)$ which weights the control value  $u_i(t)$. It is intuitive that to fulfill the purpose, the function $\gamma(t)$ must be ``comparable" with $\eps(t)$. This is achieved assuming that there exists $c>0$ such that 
\be\label{eq:no-zeno-cond}
\dst\frac{\eps(t)}{4\deg_i\gamma(t)} \ge c \quad \forall i\in I, \:\forall t\ge0.
\ee
We can now state the following result.

\begin{theorem}[Asymptotical consensus]\label{thm:asymptotical}
Let $x(\cdot)$ be the solution to~\eqref{eq:modelloB-cont}-\eqref{eq:modelloB-disc} under condition~\eqref{eq:no-zeno-cond}. 
Then, for every initial condition $\bar x\in \real^n$ there exists $\beta\in\real$ such that
$\lim_{t\to\infty}x_i(t)=\beta$ for all $i\in I$, if and only if $\int_0^{+\infty}\gamma(s)ds$ is divergent.
\end{theorem}
\begin{figure}
\psfrag{x}[][][1][-90]{$x$}  
\psfrag{V}[][][1][-90]{$V$}  
\psfrag{Time}{$t$} 
\includegraphics[width=\columnwidth]{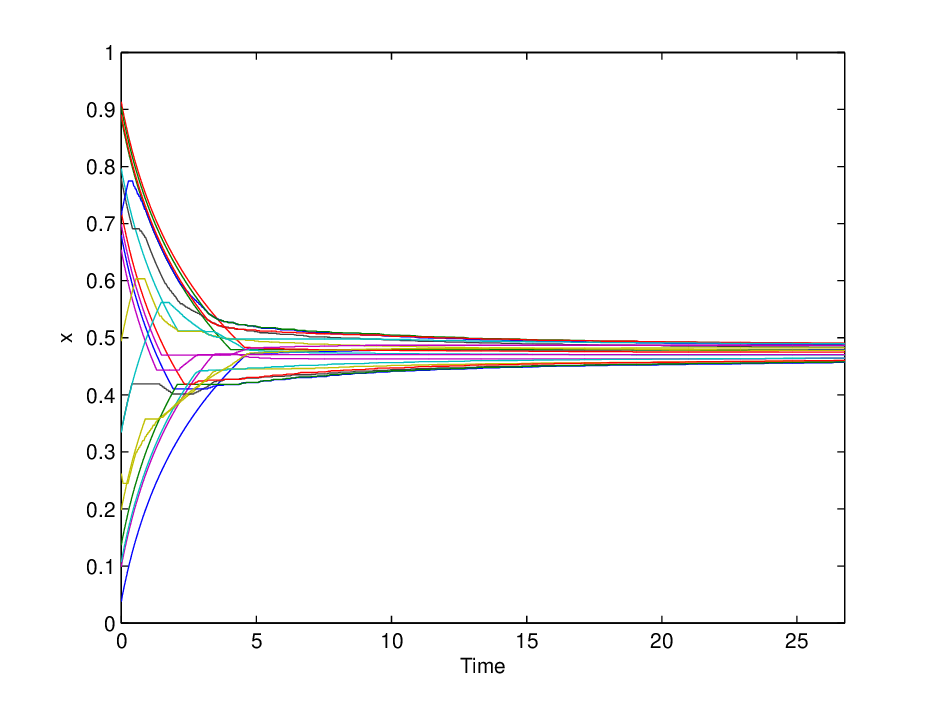}
\includegraphics[width=\columnwidth]{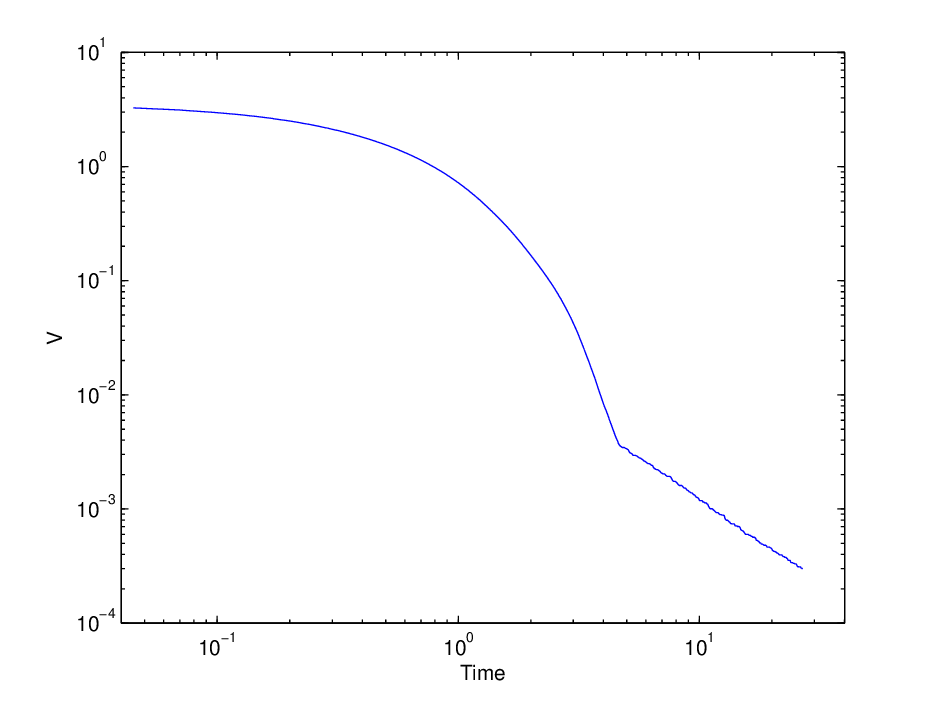}
\caption{
A sample evolution of~\eqref{eq:modelloB-cont}-\eqref{eq:modelloB-disc} starting from the same initial condition and on the same graph as Fig.~\ref{fig:simul-modelloA}. 
Top plot shows the state $x$, bottom plot shows the Lyapunov function $V$ on a logarithmic scale. 
Simulation assumes $\eps(t)=\frac{0.05}{1+t}$, $\gamma(t)=\frac{0.25}{1+t}$.
}
\label{fig:simul-modelloB}
\end{figure}
\begin{figure}
{
\psfrag{x}[][][1][-90]{$x$} 
\psfrag{data1}[][][1][0]{$x_1$} 
\psfrag{data2}[][][1][0]{$x_2$} 
\psfrag{data3}[][][1][0]{$x_3$} 
\psfrag{data4}[][][1][0]{$x_4$} 
\psfrag{data5}[][][1][0]{$x_5$} 
\psfrag{u1}[][][1][-90]{ $u_1$} 
\psfrag{u2}[][][1][-90]{ $u_2$} 
\psfrag{u3}[][][1][-90]{ $u_3$} 
\psfrag{u4}[][][1][-90]{$u_4$} 
\psfrag{u5}[][][1][-90]{$u_5$} 
\psfrag{Time}{$t$} 
\includegraphics[width=\columnwidth]{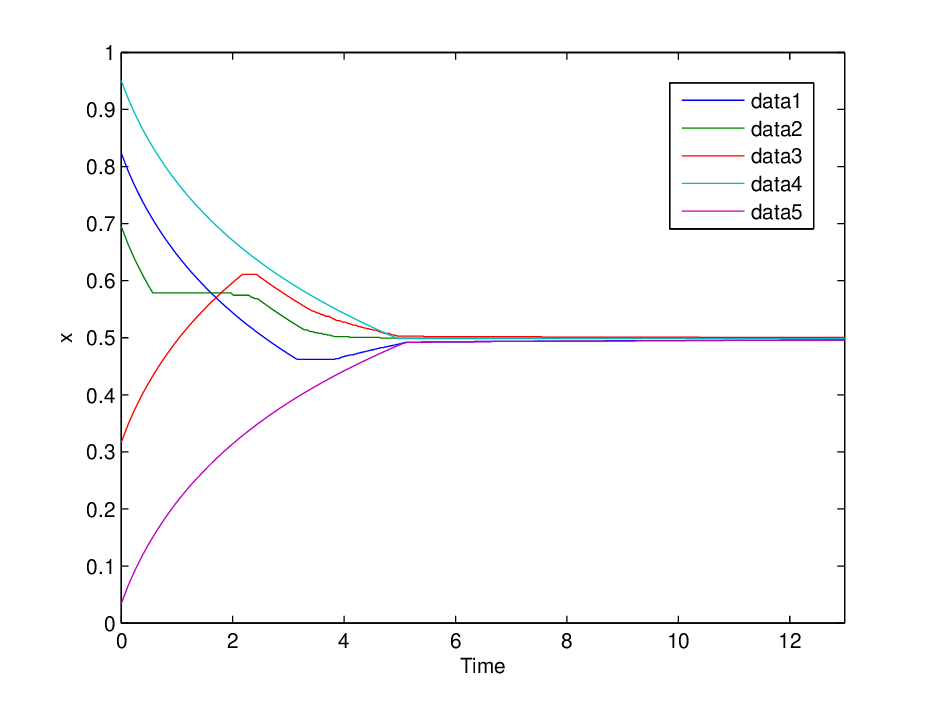}
\includegraphics[width=\columnwidth]{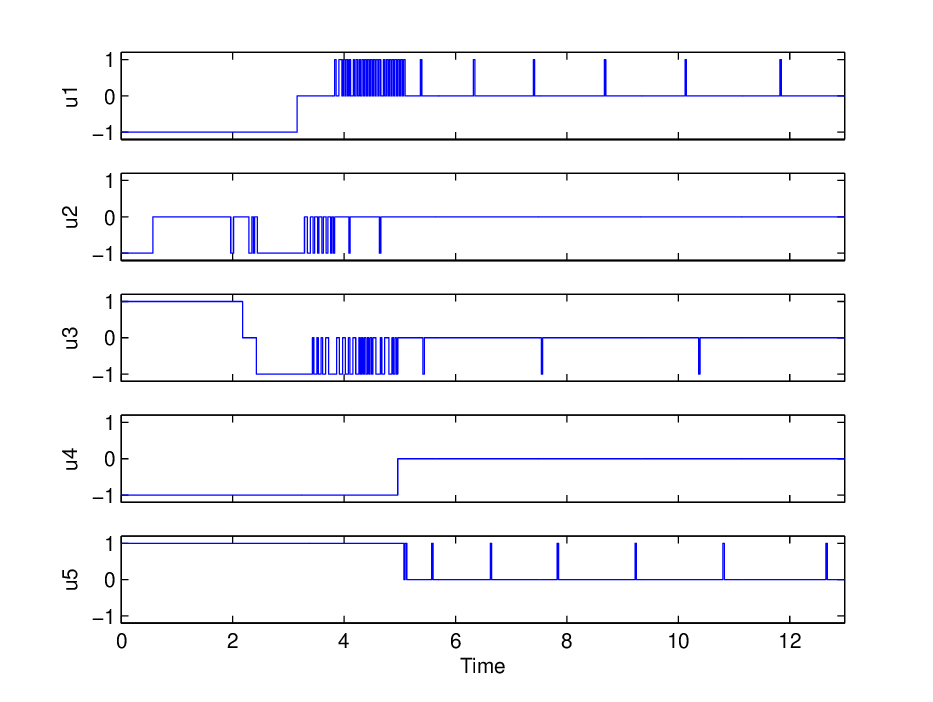}
}
%
\caption{Sample evolutions of states $x$ and corresponding controls $u$ in~\eqref{eq:modelloB-cont}-\eqref{eq:modelloB-disc} on a ring with $n=5$ nodes, $\eps(t)=\frac{0.05}{1+t}$, $\gamma(t)=\frac{0.25}{1+t}$. The guaranteed minimal inter-switching time of the controllers is $0.025$ units of time. }
\label{fig:simul-modelloB-controls}
\end{figure}
Before the proof, we discuss some illustrative simulation results, collected in Figures~\ref{fig:simul-modelloB}-\ref{fig:simul-modelloB-controls}. 
Simulations show that the evolution of~\eqref{eq:modelloB-cont}-\eqref{eq:modelloB-disc} can be qualitatively divided into two phases. During a first phase of {\em fast} convergence, $\eps$ plays a little role and the behavior of $x$ is reminiscent of Fig.~\ref{fig:simul-modelloA}.
The first phase lasts until the states become close enough to consensus for $\eps(t)$ to be comparable with the differences between the $x_i$s. During the second phase, we can observe that the control actions (\ie, the intervals of non-zero control) become sporadic and convergence depends on the decrease of $\eps$: its speed is thus {\em slow}.
Indeed, we are assuming that $\eps(t)$ has divergent integral, so that it may not decrease faster than $1/t$. 
We want to stress that this technical condition is not due to a limitation of our analysis, but is inherent to the system. Indeed, assumption~\eqref{eq:no-zeno-cond} relates $\eps$ and $\gamma$, so that $\gamma(t)$ may not be larger than (a constant times) $\eps(t)$. In turn, if $\gamma(t)$  had bounded integral, the control would not be large enough to stabilize the system to practical consensus from an arbitrary initial condition.

\begin{IEEEproof}[Proof of Theorem~\ref{thm:asymptotical}]
First of all, we note that by~\eqref{eq:no-zeno-cond}, there exists a unique solution for every initial condition. In fact, observe that 
\be\label{eq:st-asympt}
t_{k+1}^i=t_k^i +
\left\{\ba{ccc}
\dst\frac{|\ave_i(t_k^i )|}{4 d_{i}\gamma (t_k^i)} & {\rm if}  &
|\ave_i(t_k^i )|\ge \eps(t_k^i)\\[4mm]
\dst\frac{\eps(t_k^i)}{4 d_{i}\gamma (t_k^i)} & {\rm if} & |\ave_i(t_k^i)|< \eps(t_k^i),
\ea
\right.
\ee
and hence $t_{k+1}^i-t_k^i\ge c$ for all $i\in \{1,2,\ldots, n\}$ and each $k\in \Z_{\ge 0}$. 

Next, we show by an example that $\int_0^{+\infty} \gamma(s)ds=+\infty$ is necessary. Assume by contradiction that $\int_0^{+\infty} \gamma(s)ds=K\in \real$ and consider a system with $n=2$ such that $\bar x_1=K+1$ and $\bar x_2=-(K+1)$. 
Note that $|u_i(t)|\le 1$ for $i=1,2$ and $t\ge0$. Then for all $t\ge0$ we have $\dot x_1(t)\ge - \gamma(t)$ and $x_1(t)\ge x_1(0)-\int_0^t \gamma(s) ds\ge x_1(0)-K=1$: by an analogous reasoning about $x_2$, we obtain that $x_1(t)-x_2(t)\ge 2$ for all $t\ge0$, contradicting convergence.

Next, by a Lyapunov argument we show that $\int_0^{+\infty} \gamma(s)ds=+\infty$ is also sufficient. 
To this goal, we introduce the following notation. For every $t\ge0$ and $i\in I$, we consider the sequence of switching times $t^i_k$ and let 
\be\label{eq:k-star-def}k^\star_i(t):=\max\setdef{k\ge 0}{t^i_k\le t}\ee
Moreover, 
for every $t\ge0$, we let $V(t)=\frac{1}{2}{x^T(t) L x(t)}$, and we note that $V(t)\ge 0$ and 
\begin{align*}
V(t)\le& N \max_{i\in I} |x_i(t)| \max_{i\in I} |\ave_i(t)|\\
\le &N \max_{i\in I} |x_i(0)| \max_{i\in I} |\ave_i(t)|
\end{align*}
since $\max_{i\in I}x_i(t)$ (resp. $\min_{i\in I}x_i(t)$) is non-increasing in time (resp., non-decreasing). Indeed, 
let $m(t)=\max_{i\in I} x_i(t)$ and $\mu(t)=\argmax_{i\in I} x_i(t)$ and note that also Protocol~B ensures that at all times $\sign(\ave_i(t))=\sign(\ave_i(t_{k^\star_i(t)}^i))$ for every $i\in I$. In particular the latter is true for $i=\mu(t)$. 
Since $x_{\mu(t)}(t)=m(t)$, then $\ave_{\mu(t)}(t)\le 0$ and therefore $\ave_{\mu(t)}(t_{k^\star_{\mu(t)}(t)}^{\mu(t)})\le 0$. It follows by (\ref{eq:modelloB-disc}) that $u_{\mu(t)}(t)\le 0$, which implies that $\dot m(t)=\dot x_{\mu(t)}(t)\le 0$. Hence, during continuous evolution $m(t)$ cannot increase and during discrete transitions it remains constant. This shows the non-increasing monotonicity of $\max_{i\in I}x_i(t)$. Similarly, one proves the non-decreasing monotonicity of $\min_{i\in I}x_i(t)$. 


Then, we need to consider the evolution of $\dot V(t)$ along the solution of the system. 
Similarly to the proof of Theorem~\ref{thm:modelloA-convergence}, we have $\dot V(s)=x^T(s) L \gamma(s) u(s),$
and we may deduce that for all $s\ge0$, 
\[ 
\dot V(s)\le -\gamma(s)\dst\sum_{i:|\ave_{i} (t^i_{k^\star_i(s)})|\ge \eps(t^i_{k^\star_i(s)})} \frac{|\ave_i(t^i_{k^\star_i(s)})|}{2},
\]
which in particular implies that $V(t)$ is non-increasing.
It is also useful to notice that for all $t\in [t^i_{k^\star_i(t)}, t^i_{k^\star_i(t)+1}),$ 
if $|\ave_i(t^i_{k^\star_i(t)})|< \eps(t^i_{k^\star_i(t)})$, then
\begin{align*}
|\ave_i(t)| \le
 & |\ave_i(t^i_{k^\star_i(t)})|+\deg_i \dst\int_{t^i_{k^\star_i(t)}}^ t \gamma(s) ds\\
 \le & |\ave_i(t^i_{k^\star_i(t)})|+\deg_i \gamma(t^i_{k^\star_i(t)})
(t-t^i_{k^\star_i(t)})\\ 
\le & \dst\frac{5}{4}\eps(t^i_{k^\star_i(t)}). 
\end{align*}
Similarly, if $|\ave_i(t^i_{k^\star_i(t)})|\ge  \eps(t^i_{k^\star_i(t)})$, then 
\begin{align*}
|\ave_i(t)| &\le|\ave_i(t^i_{k^\star_i(t)})|+2\deg_i \dst\int_{t^i_{k^\star_i(t)}}^ t \gamma(s)\\ &\le 
\frac{3}{2} |\ave(t^i_{k^\star_i(t)})|.
\end{align*}

This inequality implies that for all $s\ge0$,
\begin{align}\label{iit}
V(s)\le N \max_{i\in I} |x_i(0)| \frac32 \max_{i\in I} \max\{ |\ave_i(t^i_{k^\star_i(s)})|, \eps(t^i_{k^\star_i(s)})\}.
\end{align}

Next, we claim that for all $\delta>0$ and $T>0$, there exists $t\ge
T$ such that $ |\ave_i(t^i_{k^\star_i(t)})|<\delta$ for all $i\in I$.
Indeed, by contradiction there would exist $\delta>0$ and $T>0$ such
that for all $t\ge T$, $
|\ave_i(t^i_{k^\star_i(t)})|\ge\delta$ for some $i\in I$, implying $\dot V(t)\le -\frac\delta2\gamma(t)$ and thus contradicting the positivity of $V(t).$
%

Since (i) the above claim holds true and (ii) $\eps(t)$ converges to zero as $t$ goes to infinity, 
we argue that for every $\delta'>0$ it is possible to choose $T'\ge0$ such that 
$\max_{i\in I} \max\{ |\ave_i(t^i_{k^\star_i(T')})|, \eps(t^i_{k^\star_i(T')})\}<\delta'$. To complete the argument, we claim that for any $\lambda >0$, there exists $T_\lambda$ such that $V(t)< \lambda$ for all $t>T_\lambda$. 
To show the latter, choose $\delta'$ such that  $ N \max_{i\in I} |x_i(0)| \frac32 \delta'\le \lambda$, and fix $T'$ accordingly. Then by (\ref{iit}), $V(T')\le \lambda$. As we have shown that $V(t)$ is monotone non-increasing, it is also true that $V(t)<\lambda$ for all $t>T'$, which  proves the claim with $T_\lambda=T'$. Hence $V(t)$ goes to zero as $t$ goes to infinity.
This fact in turn implies that $|x_i-x_j|\to 0$ as $t\to+\infty.$ 

Finally, we need to show that each trajectory converges to one point in the subspace satisfying the above condition. To this goal, we need to study the trajectories. Recall from the first part of the proof that 
$m(t)=\max_{i\in I} x_i(t)$ 
is monotonically non-increasing
and that $m(t)\ge \min_{i\in I} x_i(0)$. Hence, $\lim_{t\to+\infty}m(t)$ exists finite, which together with the result above implies convergence of all $x_i$'s to a common limit point.
\end{IEEEproof}

\subsection{Non-uniform weight functions}

According to Protocol B, the agents are required to agree on the functions $\gamma, \eps$. 
Yet this assumption is not necessary, provided that the protocol is suitably modified.

Let $\gamma_i, \eps_i$, for $i\in I$, be positive functions such that (i) they converge to zero as time diverges, (ii) their integrals on $(0,+\infty)$ are infinite, and (iii) the inequality
\be\label{eq:no-zeno-cond:2}
\dst\frac{\eps_i(t)}{2\sum_{j\in \neigh{i}}(\gamma_j(t)+\gamma_i(t))} \ge c_i \quad \forall i\in I, \:\forall t\ge0
\ee
holds for some positive numbers $c_i$. 
Let   the control law in the first equation of  (\ref{eq:modelloB-cont}) be replaced by
\be\label{cem}
\dot x_i=\gamma_i u_i.
\ee
 As for the discrete evolution,  replace the update law for $u_i$ with 
\be
u_i(t^+)=
\begin{cases}
\sign_{\eps_i(t)}\left(\ave_i(t)\right) \quad \text{if}\: i\in \I(\theta,t)\\
u_i(t) \quad \text{otherwise}
\end{cases}
\ee
and the one for
 $\theta_i$ in  (\ref{eq:modelloB-disc}) with the following:
\be\label{thetam}
 \theta_i(t^+)=\begin{cases}
 \theta_{i}(t)+ \dst \frac{\bar f_i(x(t))}{\sum_{j\in \neigh{i}}(\gamma_j(t)+\gamma_i(t))} \quad \text{if}\: i\in \I(\theta,t)\\
\theta_i(t) \quad \text{otherwise},
\end{cases}
\ee
where  
\begin{equation}\label{bar_f_i}
\bar f_i(x)=\begin{cases}
\dst\frac12{| \dst\sum_{j\in\neigh{i}}(x_j-x_i)|}& \quad \text{if}\: |\dst\sum_{j\in\neigh{i}}(x_j-x_i)|\ge\eps_i\\
\dst\frac{\eps}{2} &\quad \text{otherwise}.
\end{cases}
\end{equation}
The following result shows that the requirement on each agent using the same functions $\gamma, \eps$ can be relaxed provided that the agents locally exchange information regarding the functions~$\gamma_i$:
\begin{proposition}
Let $x(\cdot)$ be the solution to~\eqref{eq:modelloB-cont}-\eqref{eq:modelloB-disc}, modified according to (\ref{cem}),
(\ref{thetam}) and (\ref{bar_f_i}), and let  condition~\eqref{eq:no-zeno-cond:2} hold. Then there exists $\beta\in\real$ such that
$\lim_{t\to\infty}x_i(t)=\beta$ for all $i\in I.$
\end{proposition}

\begin{IEEEproof}
The arguments follow  the lines of the proof of Theorem \ref{thm:asymptotical}
and hence some details  are omitted.\\ 
In view of (\ref{thetam}) and (\ref{bar_f_i}), it is clear  that 
\be\label{st.mod}
t_{k+1}^i=t_k^i +
\left\{\ba{ccc}
\dst\frac{|\ave_i(x(t_k^i ))|}{2\Gamma_i(t_k^i )} & {\rm if}  &
|\ave_i(x(t_k^i ))|\ge \eps_i(t_k^i)\\[4mm]
\dst\frac{\eps_i(t_k^i)}{2\Gamma_i(t_k^i )} & {\rm if} & |\ave_i(x(t_k^i ))|< \eps_i(t_k^i),
\ea
\right.
\ee
with $\Gamma_i(s):=\sum_{j\in \neigh{i}}(\gamma_j(s)+\gamma_i(s))$, 
and by \eqref{eq:no-zeno-cond:2}
\[
t_{k+1}^i-t_k^i\ge \frac{\eps_i(t_k^i)}{2 \Gamma_i(t^i_k)}\ge c_i,  
\]
\ie~the times at which each agent polls the neighbors for information are separated by at least $c_i$ units of time. Hence, solutions of the system are well-defined and unique. 
The Lyapunov function $V(t)=\frac{1}{2}{x^T(t) L x(t)}$ computed along the solutions of the system satisfies 
\begin{align*}
&\dot V(t)=x^T(t) L\,{\rm diag}(\gamma_1(t),\ldots, \gamma_n(t)) \,u(t)\\
& = 
-\!\!\!\!\!\!
\dst\sum_{i:|\ave_{i} (t_k^i)|\ge {\eps_i(t_k^i)}} \!\!\!\!
\!\!\!\!\!\!
\gamma_i(t)\sign_{\eps_i(t_k^i)}\left({\ave_{i} (t_k^i)}\right)
\dst\sum_{j\in \neigh{i}} (x_j(t)-x_i(t)),
\end{align*}
where to make the notation compact we are using $t_k^i$ instead of $t^i_{k_i^*(t)}$ as defined in~\eqref{eq:k-star-def}. 
Observe that during continuous evolution
\[
\left|\frac{d}{dt} \ave_i(t)\right| \le 
\Gamma_i(t).
\]
In view of the last inequality and of \eqref{st.mod}, if $|\ave_{i} (t_k^i)|\ge \eps_i(t_k^i)$, then the bound 
\begin{align*}
\dst\frac{|\ave_i(t_k^i)|}{2}
\le 
\left|
\ave_i(t)
\right|
\end{align*}
holds for $t\in [t_{k}^i, t_{k+1}^i]$, and $\ave_i(t)$
preserves the sign.  Hence, if $|\ave_{i} (t_k^i)|\ge \eps_i(t_k^i)$, then 
\begin{align*}
\!\!\!\!\!\!
\dst\sum_{i:|\ave_{i} (t_k^i)|\ge {\eps_i(t_k^i)}} \!\!\!\!
\!\!\!\!\!\!
\gamma_i(t) \ave_i(t) &
\sign_{\eps_i(t_k^i)}\left({\ave_{i} (t_k^i)}\right)=
\\
=&
\!\!\!\!\!\!
\dst\sum_{i:|\ave_{i} (t_k^i)|\ge {\eps_i(t_k^i)}} \!\!\!\!
\!\!\!\!\!\!
\gamma_i(t)\left|
\ave_i(t)
\right|
\\ \ge &
\dst\sum_{i:|\ave_{i} (t_k^i)|\ge \eps_i(t_k^i)} \gamma_i(t)\frac{|\ave_i(t_k^i)|}{2},
%
%
\end{align*}
and we conclude that 
\be\label{wct}
\dot V(t)\le -\dst \!\!\!\!\sum_{i:|\ave_{i} (t_k^i)|\ge \eps_i(t_k^i)}
\!\!\!\! \gamma_i(t)\frac{|\ave_i(t_k^i)|}{2},
\ee
that is $V(t)$ is non-increasing. Furthermore, for all 
$t\in [t^i_{k}, t^i_{k+1}),$ and all $i\in I$, 
\begin{align*}
|\ave_i(t)| &\le 
\frac{3}{2} \max\{|\ave(t^i_{k})|, |\eps_i(t^i_{k})|\}.
\end{align*}

Exploiting the preservation of the sign of $\ave_i(t)$ during continuous evolution, we can again show that $\max_{i\in I} x_i(t)$ and 
 $\min_{i\in I} x_i(t)$ are monotone non-increasing and non-decreasing functions, respectively. This property and the last inequality above show that
 for all $t\ge0$,
$$
V(t)\le N \max_{i\in I} |x_i(0)| \frac32 \max_{i\in I} \max\{ |\ave_i(t^i_{k})|, \eps_i(t^i_{k})\}.
$$
Inequality (\ref{wct}) and the non-negativity of $V$ implies that $|\ave_i(t^i_{k})|$ must converge to zero as $t\to +\infty$. This fact and the monotonicity of $V(t)$ give convergence to zero of $V$, which in turn implies convergence of all states to a consensus.
\end{IEEEproof}


\section{Independent polling of neighbors}\label{sec.independent}

In Protocol A,  each time an agents polls its neighbors, it polls all of them simultaneously. However,  it is possible to design a similar protocol so that each agent collects information from a neighbor independently of its other neighbors. This modification leads to similar convergence results, as we shall see in what follows.

Let us adopt a new set of state variables $(x,u,\theta)$, which take value in the state space $\R^n\times \R^{d}\times \R^d$, where $d$ is the sum of the neighbors of all the agents, namely $d=\sum_{i=1}^n d_i$. 
The continuous evolution of the system obeys the equations
\be 
\label{cip}
\begin{cases}
\dot x_i=\sum_{j\in\neigh{i}}u_i^j\\
\dot u_i^j=0\\
\dot \theta_i^j=-1
\end{cases}\ee
where $i\in I$ and $j\in \neigh{i}$. The system satisfies the differential equation above for all $t$ except for those values of the time at which  the set  $${\cal J}(\theta,t)=\setdef{(i,j)\in I\times I}{j\in \neigh{i} \:\text{and}\:\theta_i^j(t)=0}$$ is non-empty. At these times a discrete transition occurs, which is governed by the following discrete update:
\be 
\label{dip}
\begin{cases} x_i(t^+)=x_i(t) \quad \forall i\in I\\
u_i^j(t^+)=
\begin{cases}
\sign_{\eps}\!\big(x_j(t)-x_i(t)\big) \quad \text{if}\: (i,j)\in {\cal J}(\theta,t)\\
u_i^j(t) \quad \text{otherwise}
\end{cases}\\
 \theta_i^j(t^+)=\begin{cases}
f_i^j(x(t)) \quad \text{if}\: (i,j)\in {\cal J}(\theta,t)\\
\theta_i^j(t) \quad \text{otherwise}
\end{cases}\\
\end{cases}
\ee
where for every $i\in I$ and $j\in \neigh{i}$, the map $\map{f_i^j}{\real^n}{\realpositive}$ is defined by 
\be
\label{fdip}
f_i^j(x)=\begin{cases}
\dst\frac{|x_j-x_i|}{2 (\deg_i+\deg_j)}& \quad \text{if}\: |x_j-x_i|\ge\eps\\[2mm]
\dst\frac{\eps}{2(\deg_i+\deg_j)} &\quad \text{otherwise}.
\end{cases}
\ee
We denote the $k$th time $t$ at which $(i,j)\in {\cal J}(\theta,t)$ by $t_k^{ij}$.

The new protocol can be described as follows:
\smallskip
\myrule
\vspace{.75\smallskipamount}
\noindent\hfill\textbf{Protocol C}\hfill\vspace{.75\smallskipamount}
\myrule\vspace{.75\smallskipamount}
\begin{algorithmic}[1]
\STATE {\bf initialization:} for all $i\in I$, for all $j\in \neigh{i}$, set $\theta_i^j=0$, $u_i^j(0)\in \{-1,0,+1\}$, and $u_i(0)=\sum_{j\in \neigh{i}} u_i^j(0)$;
\FORALL{$i\in I$}
\FORALL{$j\in \neigh{i}$}
\WHILE{$\theta_i^j(t)>0$} 
\STATE $i$ applies the control $u_i(t)=\sum_{j\in \neigh{i}} u_i^j(t)$;
\ENDWHILE
\IF{$\theta_i^j(t)=0$}
\STATE $i$ polls $j$ and collects the information $x_j(t)-x_i(t)$;
\STATE $i$ updates $\theta_i^j(t^+)=f_i^j(x(t));$
\STATE $i$ updates $u_i^j(t^+)=\sign_{\eps}\!\big(x_j(t)-x_i(t)\big)$;
\ENDIF
\ENDFOR
\ENDFOR
\end{algorithmic}
\vspace{.5\smallskipamount}\myrule\smallskip
\medskip

In contrast with Protocol A, we note that in this case the control applied by each agent is a sum of ternary controls.
Moreover, it holds for all $\{i,j\}\in E$ that $\theta_i^j(t)=\theta_j^i(t)$ and $u_i^j(t)=-u_j^i(t)$ for all $t\ge0$. This edge synchrony is essential in the analysis which follows, and points to the fact that Protocol~C is actually an {\em edge-based} algorithm, although in the proposed implementation the active entities are the agents, \ie, the nodes. 
On this respect, this feature is reminiscent of several pairwise ``gossip'' approaches which have appeared in the literature\footnote{References include randomized~\cite{SB-AG-BP-DS:06} and deterministic~\cite{JL-SM-ASM-BDOA-CY:11} approaches, with applications ranging from signal processing~\cite{AC-FF-LS-SZ:10} to optimal deployment of robotic networks~\cite{FB-RC-PF:08u}.}: we might indeed term Protocol~C a ``self-triggered gossip algorithm''.

The following convergence result holds:
\begin{theorem}[Practical consensus]\label{thm:modelloA'-convergence}
For every initial condition $\bar x$, let $x(t)$ be the solution to~\eqref{cip}-\eqref{dip} such that $x(0)=\bar x$. 
Then $x(t)$ converges in finite time to a point $x^*$ belonging to the 
set $$\E'=
\setdef{x\in \R^n}{| x_j-x_i |< \eps \;\forall\,\{i,j\}\in E}.
$$
Moreover, defined $T'=\inf\setdef{t\ge0}{x(t)\in \E'}$ and 
$C'=\max_{\{i,j\}\in E} \max\setdef{k}{t_k^{ij}\le T'}$, $T'$ and $C'$ satisfy the same bounds  as $T,C$  in Proposition \ref{prop:costs}. 
\end{theorem}
\smallskip
We note that if $x\in {\cal E}'$, then for each pair of agents $i,j$, the distance $|x_i-x_j|$ is strictly smaller than $\eps$ times the diameter of the network.
We then argue that with the new protocol the solution converges in finite time to a set which can be more explicitly characterized 
than the set obtained with Protocol~A, while time and communication costs are not larger than those of Protocol~A. This good performance is achieved at the price of employing $d_i$ time variables $\theta_i^j$ and controls $u^j_i$ per agent, instead of a single one as in Protocol~A. 

%

\begin{IEEEproof}
[Proof of Theorem~\ref{thm:modelloA'-convergence}]
In this proof we adopt the Lyapunov function 
$V(x)=\frac{1}{2} x^T x$. 
For a given $t$, let $t_k^{ij}=\max \{t_\ell^{ij}: t_\ell^{ij}<t, \ell\in \mathbb{N}\}$. 
Along the solution of (\ref{cip}), the function satisfies
\begin{align*}
 \dot V(t)=&  \sum_{i=1}^N x_i(t) \dot x_i(t)\\
 =&  \sum_{i=1}^N x_i(t) \sum_ {j\in \neigh{i}} u_i^j(t)\\
 =&  \sum_{i=1}^N x_i(t) \sum_{j\in \neigh{i}} \sign_\eps(x_j(t_k^{ij})-x_i(t_k^{ij}))\\
=&  - \sum_{\{i,j\}\in E}(x_j(t)-x_i(t)) \sign_\eps(x_j(t_k^{ij})-x_i(t_k^{ij})). 
\end{align*}
During the continuous evolution $|\dot x_j(t)- \dot x_i(t)| \le d_i+d_j$, and at the jumps $x_i(t)-x_j(t)$ does not change its value. This implies that $x_i(t)-x_j(t)$ cannot differ from $x_i(t_k^{ij})-x_j(t_k^{ij})$ in absolute value for more than $(d_i+d_j)(t-t_k^{ij})$. Exploiting this fact, if  $|x_i(t_k^{ij})-x_j(t_k^{ij})|\ge\eps$, then by~\eqref{fdip} for all $t\in [t_k^{ij}, t_{k+1}^{ij}]$, we have 
$$
|x_i(t)-x_j(t)|\ge 
\frac{|x_i(t_k^{ij})-x_j(t_k^{ij})|}{2}
$$
and
$\sign_\eps(x_i(t)-x_j(t))=\sign_\eps(x_i(t_k^{ij})-x_j(t_k^{ij}))$.
Hence 
\[\ba{rcl}
\dot V(t) 
           &=& - \dst \sum_{\{i,j\}\in E:|x_i(t_k^{ij})-x_j(t_k^{ij})|\ge \eps} |x_i(t)-x_j(t))| \\
  &\le& - \dst \sum_{\{i,j\}\in E:|x_i(t_k^{ij})-x_j(t_k^{ij})|\ge \eps} \dst\frac{|x_i(t_k^{ij})-x_j(t_k^{ij})|}{2}.
           \ea\]
This implies that there exists a finite time $T$ such that, for all $t\ge T$,  $|x_i(t_k^{ij})-x_j(t_k^{ij})|<\eps$ for all $(i,j) \in E$, because if this were not true then there would  exist $\{i,j\}\in E$ and an infinite subsequence $t_{k'}^{ij}$ of the sequence of switching times $t_k^{ij}$ such that  $|x_i(t_{k'}^{ij})-x_j(t_{k'}^{ij})|\ge \eps$, which would contradict the positiveness of $V(t)$.\\
Hence for $t\ge T$,  $|x_i(t_k^{ij})-x_j(t_k^{ij})|<\eps$ for all $(i,j) \in E$. Moreover, if $t\ge \max_{\{i,j\}\in E}t_{k+1}^{ij}$, $u_i^j(t)=0$, the state stops evolving and satisfies $|x_i(t)-x_j(t)|<\eps$, that is the first part of the thesis.

As far as the second part of thesis is concerned, similarly to the proof of Proposition~\ref{prop:costs}, we observe that if for some $t$, $u(t)=0$, then either $u_i^j(t_{k+1}^{ij})\ne 0$ for some $i\in I$ and some $j\in \neigh{i}$ (where  $t_{k+1}^{ij}$ denotes the smallest switching time larger than $t$ at which agents $i,j$ update their variables), or $u(t')=0$ for all  $t'\ge 0$. In the latter case, the state has already reached the set ${\cal E}'$. Since we are interested to characterize the time $T'$ by which convergence is achieved, we focus on the former case. Then we see that $\dot V(t)=0$ for at most $\frac{\eps}{4}$ units of time (the maximal length of an interval of time over which $u=0$ before the state has reached ${\cal E}'$) and that the interval must be followed by an interval of at least $\frac{\eps}{4\degmax}$ units of time over which $\dot V(t)\le -\frac{\eps}{2}$. These estimates imply for $T'$ and $C'$ the same bounds as obtained for $T$ and $C$ in Proposition~\ref{prop:costs}.
\end{IEEEproof}
\smallskip

Protocol~C can also be studied in terms of robustness: quantized communication and delays can be dealt with, as we have done in Section~\ref{sect:robustness}, as long as synchrony and symmetry are preserved at the edge level: indeed we recall that these assumptions are crucial to the protocol. 
A detailed robustness analysis, however, is left to future research.

In the rest of this section, we instead present a modification of Protocol~C leading to asymptotical consensus. While its design is largely inspired by Section~\ref{sec:asymptotical}, its analysis is partly different: hence we include a proof of convergence.
%
In order to yield asymptotical consensus, the protocol is modified as follows. The continuous evolution (\ref{cip}) is replaced by 
\be 
\label{cipa}
\begin{cases}
\dot x_i=\gamma(t)\sum_{j\in\neigh{i}}u_i^j\\
\dot u_i^j=0\\
\dot \theta_i^j=-1
\end{cases}\ee
whereas the discrete evolution (\ref{dip}) is replaced by 
\be 
\label{dipa}
\begin{cases} x_i(t^+)=x_i(t) \quad \forall i\in I\\
u_i^j(t^+)=
\begin{cases}
\sign_{\eps(t)}\left(x_j(t)-x_i(t)\right) \quad \text{if}\: (i,j)\in {\cal J}(\theta,t)\\
u_i^j(t) \quad \text{otherwise}
\end{cases}\\
 \theta_i^j(t^+)=\begin{cases}
\frac{1}{\gamma(t)} f_i^j(x(t)) \quad \text{if}\: (i,j)\in {\cal J}(\theta,t)\\
\theta_i^j(t) \quad \text{otherwise}.
\end{cases}\\
\end{cases}
\ee
where $f_i^j(x)$ is defined in (\ref{fdip}) and  
the functions $\eps(t), \gamma(t)$ are as in Section~\ref{sec:asymptotical}. 
The protocol just introduced leads to the following result:
\begin{theorem}[Asymptotical consensus]\label{thm:asymptotical-single}
Let $x(\cdot)$ be the solution to~\eqref{cipa}-\eqref{dipa} under condition~\eqref{eq:no-zeno-cond}. 
Then for every initial condition $\bar x\in \real^n$ there exists $\beta\in\real$ such that
$\lim_{t\to\infty}x_i(t)=\beta$ for all $i\in I$, provided that $\int_0^{+\infty}\gamma(s)ds$ is divergent.
\end{theorem}

\begin{IEEEproof}
As in the proof of Theorem~\ref{thm:asymptotical}, one shows the equality
\[\ba{rcl}
\dot V(t)
&=& \dst  -  \;\;\gamma(t)
 \!\!\!\!\!\! \!\!\!\! \!\!\!\!\sum_{\{i,j\}\in E: |x_j(t_k^{ij})-x_i(t_k^{ij})|\ge \eps(t_k^{ij})}\!\!\!\! \!\!\frac{|x_j(t_k^{ij})-x_i(t_k^{ij}))|}{2}. 
           \ea\]
From the latter and  the properties $\eps(t)\to 0$ and  $\int_0^{+\infty}\gamma(s)ds =+\infty$, it  follows that for each $\delta>0$, for each $T_\delta>0$, there exists $t\ge T_\delta$ such that  $ |x_j(t_{k^\star_{ij}(t)})-x_i(t_{k^\star_{ij}(t)})|< \delta$ for all $(i, j)\in E$,  where   $k^\star_{ij}(t):=\max\setdef{k\ge 0}{t^{ij}_k\le t}$. 


Consider now the function $W(x)=\max_{i\in I} x_i-\min_{i\in I} x_i$. The function $W(x(t))$ is non-increasing along the solutions to~(\ref{cipa}), (\ref{dipa}). Indeed,  Protocol C guarantees that, for all $\{i,j\}\in E$, the sign of $x_j(t_{k^\star_{ij}(t)})-x_i(t_{k^\star_{ij}(t)})$ and the sign of $x_j(t)-x_i(t)$ are the same for all $t\in [t_{k^\star_{ij}(t)}, t_{k^\star_{ij}(t)+1}]$.  
Furthermore we notice that $W(x)\le \diam(G)\cdot\max_{\{i,j\}\in E} |x_i-x_j|$. 
Bearing in mind the arguments above, we now prove that $\lim_{t\to+\infty} W(x(t))=0$. Indeed, for each $\eps'>0$, fix $\delta\le \frac{\eps'}{\diam G}$ and choose $t$ sufficiently large
that  $ |x_j(t_{k^\star_{ij}(t)})-x_i(t_{k^\star_{ij}(t)})|< \delta$ for all $\{i,j\}\in E$ --the existence of such $t$ has been discussed  in the first part of the proof. 
Then 
$W(x(t_{k^\star_{ij}(t)}))\le \diam(G)\max_{\{i,j\}\in E} |x_j(t_{k^\star_{ij}(t)})-x_i(t_{k^\star_{ij}(t)})|<\eps'$. 
Since $W(x(t))$ is non-increasing then $W(x(t))\le W(x(t_{k^\star_{ij}(t)}))<\eps'$ for all $t\ge t_{k^\star_{ij}(t)}$. 
Hence we have shown that for any $\eps'>0$, there exists a time 
$T_{\eps'}:=t_{k^\star_{ij}}(t)$ such that $W(x(t))<\eps'$ for all $t\ge T_{\eps'}$, which proves  $\lim_{t\to+\infty} W(x(t))=0$. By definition of $W$ the  thesis follows.   
\end{IEEEproof}

\section{Conclusions}
In this paper we have addressed the problem of achieving consensus in the scenario in which agents collect information from the neighbors only at times which are designed iteratively and independently by each agent  on the basis of its current local  measurements, a process which following existing literature can be termed self-triggered information collection.
Compared with existing results, our approach presents a number of remarkable features. Based on the use of relative measurements only, our self triggered control policy achieves practical consensus with a guaranteed minimal inter-sampling time which can be freely tuned by the designer: remarkably, no global information on the graph topology is required for either designing or running the algorithm.
The approach lends itself to an expressive characterization of the tradeoff between controller accuracy and communications costs.  
We have also shown that our algorithm is inherently robust to uncertainties commonly found in networked systems, with the margin of robustness being adjustable via appropriate tuning of certain design parameters. 
To achieve asymptotic consensus, we have proposed a modification of our basic self-triggered control scheme. 
Finally we have identified a third communication protocol, in which agents communicate in a  pairwise  fashion at times which are designed iteratively, a protocol   which  we proposed to name self-triggered gossiping algorithm. Besides its inherent deterministic nature, this gossiping-like algorithm appears to be one of the first to have been specifically devised  for continuous-time systems. 

From the methodological point of view, most of our results descend from Lyapunov-like analysis of the class of hybrid systems used to model our distributed self-triggered control schemes. 
Some research questions have been left open for future research, such as the robustness of the self-triggered algorithm which guarantees asymptotic convergence. 

Besides collateral issues, we envisage three main avenues for new research stemming from this work.
First, we recall that the ternary nature of controllers has a key role in our approach, as it provides implicit information on the dynamics, which is exploited in the computation of the sampling times. Thus, a natural extension would be to consider constrained controllers taking values in larger sets, for instance saturated controllers.
Second, further investigation and extensions of the self-triggered gossip algorithm introduced in this paper may enhance the already rich literature on gossip algorithms.  
Third, a very interesting question concerns how similar approaches perform in the case of higher dimensional systems and for more complex coordination tasks.

%
%
\bibliographystyle{ieeetran}

\begin{thebibliography}{10}
\providecommand{\url}[1]{#1}
\csname url@samestyle\endcsname
\providecommand{\newblock}{\relax}
\providecommand{\bibinfo}[2]{#2}
\providecommand{\BIBentrySTDinterwordspacing}{\spaceskip=0pt\relax}
\providecommand{\BIBentryALTinterwordstretchfactor}{4}
\providecommand{\BIBentryALTinterwordspacing}{\spaceskip=\fontdimen2\font plus
\BIBentryALTinterwordstretchfactor\fontdimen3\font minus
  \fontdimen4\font\relax}
\providecommand{\BIBforeignlanguage}[2]{{%
\expandafter\ifx\csname l@#1\endcsname\relax
\typeout{** WARNING: IEEEtran.bst: No hyphenation pattern has been}%
\typeout{** loaded for the language `#1'. Using the pattern for}%
\typeout{** the default language instead.}%
\else
\language=\csname l@#1\endcsname
\fi
#2}}
\providecommand{\BIBdecl}{\relax}
\BIBdecl

\bibitem{JC:06b}
J.~Cort{\'e}s, ``Finite-time convergent gradient flows with applications to
  network consensus,'' \emph{Automatica}, vol.~42, no.~11, pp. 1993--2000,
  2006.

\bibitem{CDP:09}
C.~{De Persis}, ``Robust stabilization of nonlinear systems by quantized and
  ternary control,'' \emph{Systems \& Control Letters}, vol.~58, no.~8, pp.
  602--608, 2009.

\bibitem{KA-BB:02}
K.~{\AA}str\"om and B.~Bernhardsson, ``Comparison of {Riemann} and {Lebesgue}
  sampling for first order stochastic systems,'' in \emph{{IEEE} Conf. on
  Decision and Control}, Las Vegas, NV, USA, 2002, pp. 2011--2016.

\bibitem{XW-MDL:09}
X.~Wang and M.~D. Lemmon, ``Event-triggering in distributed networked systems
  with data dropouts and delays,'' in \emph{Hybrid systems: computation and
  control}, ser. Lecture Notes in Computer Science, R.~Majumdar and P.~Tabuada,
  Eds.\hskip 1em plus 0.5em minus 0.4em\relax Springer, 2009, vol. 5469, pp.
  366--380.

\bibitem{MDB-SDG-AD:11}
M.~D. {Di Benedetto}, S.~{Di Gennaro}, and A.~{D'Innocenzo}, ``Digital self
  triggered robust control of nonlinear systems,'' in \emph{{IEEE} Conf. on
  Decision and Control and European Control Conference}, 2011, pp. 1674--1679.

\bibitem{HY-PJA:11a}
H.~Yu and P.~J. Antsaklis, ``Event-triggered output feedback control for
  networked control systems using passivity: Time-varying network induced
  delays,'' in \emph{{IEEE} Conf. on Decision and Control and European Control
  Conference}, Orlando, FL, USA, Dec. 2011, pp. 205--210.

\bibitem{DL-JL:12}
D.~Lehmann and J.~Lunze, ``Event-based control with communication delays and
  packet losses,'' \emph{International Journal of Control}, vol.~85, no.~5, pp.
  563--577, 2012.

\bibitem{PT:07}
P.~Tabuada, ``Event-triggered real-time scheduling of stabilizing control
  tasks,'' \emph{IEEE Transactions on Automatic Control}, vol.~52, no.~9, pp.
  1680--1685, 2007.

\bibitem{XW-MDL:11}
X.~Wang and M.~Lemmon, ``Event-triggering in distributed networked control
  systems,'' \emph{IEEE Transactions on Automatic Control}, vol.~56, no.~3, pp.
  586--601, 2011.

\bibitem{MM-PT:11}
M.~{Mazo Jr.} and P.~Tabuada, ``Decentralized event-triggered control over
  wireless sensor/actuator networks,'' \emph{IEEE Transactions on Automatic
  Control}, vol.~56, no.~10, pp. 2456--2461, 2011.

\bibitem{MM-MC:11}
M.~{Mazo Jr.} and M.~Cao, ``Decentralized event-triggered control with
  asynchronous updates,'' in \emph{{IEEE} Conf. on Decision and Control and
  European Control Conference}, 2011, pp. 2547--2552.

\bibitem{CDP-RS-FW:11}
C.~{De Persis}, R.~Sailer, and F.~Wirth, ``On a small-gain approach to
  distributed event-triggered control,'' in \emph{{IFAC} {W}orld {C}ongress},
  Milan, Italy, 2011, pp. 2401--2406.

\bibitem{RB-FA:11}
R.~Blind and F.~Allg{\"o}wer, ``Analysis of networked event-based control with
  a shared communication medium: {P}art {I} - pure {ALOHA},'' in \emph{{IFAC}
  {W}orld {C}ongress}, Milan, Italy, 2011, pp. 10\,092--10\,097.

\bibitem{FC-CDP-PF:10a}
F.~Ceragioli, C.~{De~Persis}, and P.~Frasca, ``Discontinuities and hysteresis
  in quantized average consensus,'' \emph{Automatica}, vol.~47, no.~9, pp.
  1916--1928, 2011.

\bibitem{GS-DVD-KHJ:11}
G.~Seyboth, D.~V. Dimarogonas, and K.~H. Johansson, ``Control of multi-agent
  systems via event-based communication,'' in \emph{{IFAC} {W}orld {C}ongress},
  Milan, Italy, Aug. 2011, pp. 10\,086--10\,091.

\bibitem{DVD-EF-KHJ:12}
D.~V. Dimarogonas, E.~Frazzoli, and K.~H. Johansson, ``Distributed
  event-triggered control for multi.agent systems,'' \emph{IEEE Transactions on
  Automatic Control}, vol.~57, no.~5, pp. 1291--1297, 2012.

\bibitem{CN-JC:12}
C.~Nowzari and J.~Cort\'es, ``Self-triggered coordination of robotic networks
  for optimal deployment,'' \emph{Automatica}, vol.~48, no.~6, pp. 1077--1087,
  2012.

\bibitem{SB-AG-BP-DS:06}
S.~Boyd, A.~Ghosh, B.~Prabhakar, and D.~Shah, ``Randomized gossip algorithms,''
  \emph{IEEE Transactions on Information Theory}, vol.~52, no.~6, pp.
  2508--2530, 2006.

\bibitem{JL-SM-ASM-BDOA-CY:11}
J.~Liu, S.~Mou, A.~S. Morse, B.~D.~O. Anderson, and C.~Yu, ``Deterministic
  gossiping,'' \emph{Proceedings of the IEEE}, vol.~99, no.~9, pp. 1505--1524,
  2011.

\bibitem{AC-FF-LS-SZ:10}
A.~Chiuso, F.~Fagnani, L.~Schenato, and S.~Zampieri, ``Gossip algorithms for
  simultaneous distributed estimation and classification in sensor networks,''
  \emph{IEEE Journal of Selected Topics in Signal Processing}, vol.~5, no.~4,
  pp. 691--706, 2011.

\bibitem{FB-RC-PF:08u}
F.~Bullo, R.~Carli, and P.~Frasca, ``Gossip coverage control for robotic
  networks: {D}ynamical systems on the space of partitions,'' \emph{SIAM
  Journal on Control and Optimization}, vol.~50, no.~1, pp. 419--447, 2012.

\end{thebibliography}

\end{document}